\documentclass{elsarticle}
\usepackage[margin=2cm]{geometry}

 \usepackage{graphics}
 \usepackage{graphicx}
\usepackage{caption}
\usepackage{subcaption}
\usepackage{amssymb}
\usepackage{amsmath}
\usepackage{color}

\newtheorem{definition}{Definition}

\usepackage[shortlabels]{enumitem}
\usepackage{algorithm,algpseudocode}
\usepackage{hyperref}
\hypersetup{
colorlinks=true
}
\usepackage{epstopdf}
\usepackage{multicol}

\biboptions{sort&compress}

\journal{arXiv}

\begin{document}

\begin{frontmatter}



\title{Empirical study of
periodic autoregressive models with additive noise - estimation and testing}

\author[label1]{Wojciech Żuławiński}
\author[label1]{Agnieszka Wy\l oma\'nska\corref{cor1}}
\ead{agnieszka.wylomanska@pwr.edu.pl}
\cortext[cor1]{Corresponding author.}
\address[label1]{Faculty of Pure and Applied Mathematics, Hugo Steinhaus Center, Wroclaw University of Science and Technology, Wyspianskiego 27, 50-370 Wroclaw, Poland}

\begin{abstract}
{Periodic autoregressive (PAR) time series with finite variance is considered as one of the most common models of second-order cyclostationary processes. However, in the real applications, the signals with periodic characteristics may be disturbed by additional noise related to measurement device disturbances or to other external sources.  Thus, the known estimation techniques dedicated for PAR models may be inefficient for such cases. When the variance  of the additive noise is relatively small, it can be ignored and the classical estimation techniques can be applied.  However, for extreme cases, the additive noise can have a significant influence on the estimation results. In this paper, we propose four estimation techniques for the noise-corrupted PAR models with finite variance distributions. The methodology is  based on Yule-Walker equations utilizing the autocovariance function. It can be used for any type of the finite variance additive noise. The presented simulation study clearly indicates the efficiency of the proposed techniques, also for extreme case, when the additive noise is a sum of the Gaussian additive noise and additive outliers. The proposed estimation techniques are also applied for testing if the data corresponds to noise-corrupted PAR model. This issue is strongly related to the  identification of informative component in the data in case when the model is disturbed by additive non-informative noise. The power of the test is studied for simulated data. Finally, the testing procedure is applied for two real time series describing  particulate matter concentration in the air. 
}
\end{abstract}

\end{frontmatter}


\section{Introduction}\label{intro}
P{eriodic} autoregressive (PAR) time series is considered as one of the most common  cyclostationary processes, \cite{aaw8}. The second-order cyclostationary processes (called also periodically correlated, PC) are useful for the description of the systems with the measurements having a periodically non-stationary random structure, \cite{gardner}. This behavior is ubiquitous in various areas of interests, including hydrology \cite{aaw8,hyd2}, climatology and meteorology \cite{met1,met2}, environmental engineering \cite{7362771,SARNAGLIA20102168}, electricity market \cite{CARO2020124642}, condition monitoring \cite{antoni2004cyclostationary}, and many others, see e.g, \cite{IEEE_1,IEEE_2}. 

The analyzed in this paper PAR models are considered as the natural generalization of the popular autoregressive moving average (ARMA) time series \cite{brockwelldavis}, where the constant parameters are replaced by the periodic deterministic functions. The ARMA time series are often used in various engineering applications in order to build the model of real system based on measured data \cite{Shin2003423}. However, if the analyzed system is operating in time-varying conditions, the coefficients of ARMA model are also time-varying, in particular --  periodic, see \cite{kuba12} for interesting examples in mechanical systems.

The main features characterizing the cyclostationary behavior of finite-variance distributed signals (and thus PAR time series) are the autocovariance or autocorrelation functions (denoted as ACVF and ACF, respectively).  Thus, most of the algorithms  for cyclostationary signals utilize empirical counterparts of ACVF and ACF in time and frequency domains, see e.g. \cite{antoni2007cyclic,csc2}. There are known various estimation methods dedicated for finite-variance PAR models. The most known technique is the Yule-Walker-based approach utilizing the time-dependent ACVF \cite{yule_walker_parma}. For other techniques see e.g., \cite{aaw16,Tesfaye2006IdentificationOP,other_parma1,IEEE_3}. All of the mentioned above estimation methods  assume the ideal case, namely that the analyzed signal corresponds to the "pure" model without any additional disturbances. However, in  real applications this situation seldom occurs and in practice the measured signal is always noise-corrupted.   The additional noise may be related to measurement device disturbances or to other sources influencing the observations. This problem was extensively discussed for the noise-corrupted ARMA-type models for different types of additive noise, see e.g., \cite{esfandiari,diversi1,MAHMOUDI2010157,CAYIR2021108118,IEEE_4,IEEE_5}. The characterization of noise-corrupted PARMA models is also discussed in the literature. However, the main attention is paid to the special case of the additive noise, namely additive outliers (i.e., independent identically distributed (i.i.d.) observations with large values appearing with a given probability), see e.g. \cite{parma_ao0,parma_an1,parma_ao4,parma_ao7,parma_ao8}. The  algorithms do not take into consideration the existence of the additive disturbances but utilize the robust estimators of the classical statistics (like ACVF in the Yule-Walker method) in order to reduce the impact of large observations visible in the signal.  However, the additive noise may have a much wider sense than just several outliers. The disturbance may be a random process with Gaussian or non-Gaussian distribution and the mentioned outliers as well. All these factors strongly depend on the modelled system. 

In this paper, we consider the PAR models with general finite-variance distributed additive noise and propose four estimation algorithms for their parameters. All of the introduced techniques take into account the existence of the additive noise and can be also used in the case of the additive outliers, which is a common situation   for instance in air pollution data \cite{doi:10.1080/03610926.2018.1533970,7362771,SARNAGLIA20102168,REISEN2019842}. The proposed approaches do not assume any specific distribution of the data, thus in this sense, they are universal for finite-variance case. The presented simulation study clearly indicates the justification for the inclusion of the existence of the additive noise in the estimation algorithms. The efficiency of the new techniques is compared with the classical Yule-Walker technique (without the assumption of the additive noise existence). By Monte Carlo simulations, we have demonstrated that the classical Yule-Walker estimator is biased in contrast to the new algorithms. We also discuss the case when the additive noise is a sequence of additive outliers. The received results confirm that the proposed techniques can also be used in this case and outperform the classical approach. The most practical case considered in this paper is the PAR model with additive noise being a  sum of two sequences, namely Gaussian additive noise and additive outliers. The presence of the Gaussian additive noise is very intuitive in practical situations. Most of the signal models in metrology assume the presence of additive noise related to sensor, signal discretization, data transmission, etc. The presented results for Monte Carlo simulations indicate the efficiency of the introduced techniques also in this case.

Based on the introduced estimation algorithms we also introduce a simple testing procedure for the identification if the real data corresponds to the pure PAR time series versus the noise-corrupted model. The identification of the additive noise in the data may have a significant influence on the selection of appropriate tools for its further analysis. To demonstrate the effectiveness of the testing procedure, we analyze the power of the tests for simulated data. Finally, we examine the real time series describing the particulate matter concentration
in the air for selected region in Brazil. Our results confirm the conclusions presented in \cite{doi:10.1080/03610926.2018.1533970}, where authors proposed the PAR model with additive outliers for a similar dataset. This paper is a continuation of our previous work \cite{ZULAWINSKI2023115131}, where we  introduced the methodology for selection  of  optimal  order and period for the considered noise-corrupted PAR model and the technique for the identification of its residuals properties.

The rest of the paper is organized as follows: in Section \ref{sec_par} we introduce the model. In Section \ref{methods}, we describe four new estimation techniques for the noise-corrupted PAR model estimation. In Section \ref{simul1}, we present the simulation studies related to the PAR model with Gaussian additive noise.  In Section \ref{sec_outl}, we provide the analysis of the simulated signals corresponding to the PAR model with additive outliers.  More precisely,  we consider two examples, when the additive noise is a sequence of additive outliers and the most complicated case when the additive noise is defined as the sum of the Gaussian additive noise and the additive outliers. The presented simulation studies indicate the efficiency of the proposed estimation methods also in such situations.   In Section \ref{testing_proc}, we describe the simple testing procedure for existence of additive noise and present the  simulation study to demonstrate its efficiency for various types of disturbances. Finally, in Section \ref{real}, we analyze two real datasets from air quality area and test if they correspond to the pure PAR model (versus noise-corrupted one). Last section concludes the paper.

\section{Periodic autoregressive model with additive noise}\label{sec_par}
The periodic autoregressive model with additive noise (called also noise-corrupted PAR model) is defined as follows
\begin{eqnarray}\label{model1}
Y_t=X_t+Z_t, ~t\in \mathbb{Z},
\end{eqnarray}
where $\{X_t\}$ is the periodic autoregressive  time series of order $p\in \mathbb{N}$ and period $T\in \mathbb{N}$.  Moreover, it is assumed that innovations  of the PAR($p$) model  (denoted further as $\{\xi_t\}$) and the additive noise $\{Z_t\}$ constitute sequences of independent random variables  with zero mean and constant variances denoted further as $\sigma_{\xi}^2$ for the innovations of the PAR($p$) model and  $\sigma_Z^2$ for the additive noise, respectively. Additionally, we consider the case when the sequences $\{X_t\}$ and $\{Z_t\}$ are independent. 

To examine the characteristics of the model defined in Eq. (\ref{model1}) there is a need to remind the definition of the finite variance PAR($p$) model and its properties. The PAR($p$) time series is a special case of the PARMA  model.
\begin{definition}\cite{vecchia1985periodic}
\label{def:parma1}
The sequence $\{X_t\}$ is a second-order PARMA(p,q) ($p,q\in \mathbb{N}$) model with period $T\in \mathbb{N}$ when it satisfies the following equation
\begin{eqnarray}\label{parma1}
X_t-\phi_1(t)X_{t-1}-\cdots-\phi_p(t)X_{t-p}=\xi_{t}+\theta_1(t)\xi_{t-1}+\cdots+\theta_q(t)\xi_{t-q}.
\end{eqnarray}
In Eq. (\ref{parma1}), the innovations $\{\xi_t\}$  constitutes a sample of i.i.d. random variables with mean equal to zero and variance $\sigma_{\xi}^2$.  Additionally, the scalar sequences $\{\phi_i(t),~i=1,...,p\}$, $\{\theta_j(t),~j=1,...,q\}$ are periodic in $t$ with the same period $T$.
\end{definition}
It is worth to highlighting, that Eq. (\ref{parma1}) can be expressed also by difference equation
\begin{eqnarray}
\Phi_t(B)X_t=\Theta_t(B)\xi_t, ~t\in \mathbb{Z},
\end{eqnarray}
where $B$ is the backward-shift operator and \begin{eqnarray}\label{polynomials}\Phi_t(z)=1-\phi_1(t)z-\cdots-\phi_p(t)z^p, \quad
\Theta_t(z)=1+\theta_1(t)z+\cdots+\theta_q(t)z^q\end{eqnarray}
are autoregressive (AR) and moving average (MA) polynomials, respectively. 

In the classical version,  $\{\xi_t\}$ is considered as a sequence of  Gaussian distributed random variables. In our simulation studies, we also consider this case. However, all presented below properties hold also for any finite variance distribution of the innovation series. 

When $p=0$, then the PARMA model given in Definition \ref{def:parma1} is called the PMA($q$)  while when  $q=0$,  it is called the PAR($p$) time series. In this paper, we consider the causal models. 
The PARMA model (and therefore PAR) can be considered also as the special case of ARMA time series with time-dependent parameters. This general model with finite variance innovations was considered for instance in \cite{yule_walker_parma,Peiris2,Singh,Priestey}, where the authors give the necessary and sufficient condition for the existence of the solution of considered models. The the infinite variance ($\alpha-$stable) case was discussed \cite{Peiris1} and in \cite{kokoszka} for time-constant coefficients. See also \cite{wylomanska1,wylomanska2}. Moreover, it was shown that unique solution of the time-dependent ARMA time series (in finite and infinite variance case) is expressed as a one-sided MA model, in terms of one-sided Green's functions. Using the results presented in \cite{Peiris2} one can show that if the zeros of the polynomial $\Phi_t(z)$ given in Eq. (\ref{polynomials}) are outside the unit circle for all $t\in \mathbb{Z}$, then the PAR($p$) time series defined in Eq. (\ref{parma1}) for $q=0$ has the unique solution that converges absolutely with probability one and it is given by the following form
 \begin{eqnarray}\label{sol}
 X_t=\sum_{j=-\infty}^tG_t(j)\xi_j,
 \end{eqnarray}
 where the coefficients $G_t(j)$ satisfy
 \begin{eqnarray}
 \sum_{j=-\infty}^{t}G_t(j)z^j=\frac{1}{\Phi_t(z)}
 \end{eqnarray}
 and are absolutely  summable. The coefficients $G_t(j)$ given in Eq. (\ref{sol}) are expressed by the Green's functions $g_t(j)$ associated with the AR polynomial $\Phi_t(z)$, i.e. they are solutions of the homogeneous difference equation $\Phi_t(B)\Psi_t=0$ on $\mathbb{Z}$ with initial condition $g_t(t)=1$ and $g_t(j)=0$ when $j=t+1,\cdots,t+p-1$. Using the results presented in \cite{Peiris2} one may conclude that in case when the coefficients $\phi_1(t),\cdots,\phi_p(t)$ are periodic in $t$ with the same period $T$, the coefficients $G_t(j)$ are also periodic in $t$ with the same period. We refer the readers to \cite{Peiris1,Peiris2} for more detailed discussion.

As it was mentioned, the PAR time series with finite variance innovations is one of the common model of the second-order cyclostationary processes. The cyclostationarity for finite variance time series $\{X_t\}$ denotes that its mean and autocovariance functions are periodic in $t$ with the same period $T$ \cite{hurd2007periodically}
\begin{eqnarray}
\mathbb{E}X_t=\mathbb{E}X_{t+T},\qquad
cov(X_t,X_{t+h})=cov(X_{t+T},X_{t+h+T}),~~h\in \mathbb{Z}.
\end{eqnarray}
Indeed, using the fact that the unique  solution of PAR model is given in Eq. (\ref{sol}) and the coefficients $G_t(j)$ are absolutely summable and  periodic  for each $t\in \mathbb{Z}$, one can easily show that $\mathbb{E}(X_t)=0$ and autocovariance function of $\{X_t\}$, namely $cov(X_t,X_{t+h})$ is periodic in $t$ with period $T$ for any $h\in \mathbb{Z}$. 

Using the above facts one can show that the  PAR($p$) model with the additive noise defined in Eq. (\ref{model1}) is also second-order cyclostationary, however it does not satisfy the PAR or PARMA equation. The detailed proof for $p=1$ is presented in \cite{nasza_wojtek}. In the general case, one can show that for each $t$ the time series $\{Y_t\}$ given in Eq. (\ref{model1})  satisfies the following equation 
\begin{eqnarray}\label{y1}
Y_t-\phi_1(t)Y_{t-1}-\cdots -\phi_p(t)Y_{t-p}=\xi_t+Z_t-\phi_1(t)Z_{t-1}-\cdots-\phi_p(t)Z_{t-p}.
\end{eqnarray}
In the next part of this section, we introduce the notation used in the estimation algorithms. Putting $t=nT+v$, where $n \in \mathbb{Z},\,v=1,2,\cdots,T$ one can express the PAR($p$) model in the following form
\begin{eqnarray}\label{parp}
X_{nT+v}-\phi_1(v)X_{nT+v-1}-\cdots-\phi_p(v)X_{nT+v-p} = \xi_{nT+v}.
\end{eqnarray}
Multiplying Eq. (\ref{parp}) by $X_{nT+v-1}, \cdots, X_{nT+v-p}$  for each $v=1,2,\cdots,T$ and taking the expected value of both sides of the above equations one obtains the following system of equations
\begin{eqnarray}\label{x_loyw}
\mathbf{\Gamma}^X_v \mathbf{\Phi}_v = \mathbf{\gamma}^X_v,
\end{eqnarray}
where $\mathbf{\Gamma}^X_v$ is a $p \times p$ matrix of elements defined as follows
\begin{eqnarray}\label{x_loyw_def1}
(\mathbf{\Gamma}^X_v)_{i,j} = \gamma^X(v-i,j-i),
\end{eqnarray}
for $\gamma^X(w,k) = \mathbb{E}X_{nT+w}X_{nT+w-k}$ and
\begin{eqnarray}\label{x_loyw_def2}
\mathbf{\gamma}^X_v = [\gamma^X(v,1),\cdots,\gamma^X(v,p)]',
\end{eqnarray}
\begin{eqnarray}\label{x_loyw_def3}
\mathbf{\Phi}_v = [\phi_1(v),\cdots,\phi_p(v)]'.
\end{eqnarray}
Analogously, the same procedure starting with the multiplication by $X_{nT+v}$ yields the following expression
\begin{eqnarray}\label{x_loyw0}
\gamma^X(v,0) - \mathbf{\gamma}_v^{X \prime} \mathbf{\Phi}_v = \sigma^2_\xi.
\end{eqnarray}
The systems defined by Eqs. (\ref{x_loyw}) and (\ref{x_loyw0}) are also known as low-order Yule-Walker equations. However, we can also consider high-order Yule-Walker equations which are constructed in a similar manner. The only difference is that during the first step we multiply Eq. (\ref{parp}) by $X_{nT+v-(p+1)}, \cdots, X_{nT+v-(p+s)}$, where $s$ is a desired number of equations. As a result, for each $v=1,2,\cdots,T$ we obtain the following system
\begin{eqnarray}\label{x_hoyw}
_s\tilde{\mathbf{\Gamma}}^{X}_v \mathbf{\Phi}_v =      {}_s\tilde{\mathbf{\gamma}}^{X}_v,
\end{eqnarray}
where $_s\tilde{\mathbf{\Gamma}}^{X}_v$ is a $s \times p$ matrix of elements
\begin{eqnarray}\label{x_hoyw_def1}
({}_s\tilde{\mathbf{\Gamma}}^{X}_v)_{i,j} = \gamma^X(v-j, p+i-j),
\end{eqnarray}
and:
\begin{eqnarray}\label{x_hoyw_def2}
_s\tilde{\mathbf{\gamma}}^{X}_v = [\gamma^X(v,p+1),\cdots,\gamma^X(v,p+s)]'.
\end{eqnarray}
Although the presented Yule-Walker equations were derived for the pure PAR($p$) model $\{X_t\}$, they are also useful for the noise-corrupted time series  given in Eq. (\ref{model1}). The periodic autocovariance of $\{Y_t\}$ is
\begin{eqnarray}\label{peacvf_y}
\gamma^Y(w,k) =  \mathbb{E}Y_{nT+w}Y_{nT+w-k} =\mathbb{E}(X_{nT+w}+Z_{nT+w})(X_{nT+w-k}+Z_{nT+w-k}) =
\begin{cases}
      \gamma^X(w,k) + \sigma^2_Z& \text{$k=0$}, \\
      \gamma^X(w,k) & \text{$k\neq 0$}.
    \end{cases}
\end{eqnarray}
Using the same notation as for $\{X_t\}$ time series, let us define $\mathbf{\Gamma}^Y_v$ and $\mathbf{\gamma}^Y_v$ (see corresponding Eqs. (\ref{x_loyw_def1}) and (\ref{x_loyw_def2})). Because, from Eq. (\ref{peacvf_y}), $\gamma^Y(v,k)$ and $\gamma^X(v,k)$ are not equal only when $k=0$, thus we have $\mathbf{\gamma}^Y_v=\mathbf{\gamma}^X_v$. Moreover, the following relation holds
\begin{eqnarray}\label{acvf_x_y_1}
\mathbf{\Gamma}^Y_v = \mathbf{\Gamma}^X_v + \sigma^2_Z \mathbf{I}_p.
\end{eqnarray}
Hence, for each $v=1,2,\cdots,T$, one can rewrite the low-order Yule-Walker equations given in  (\ref{x_loyw}) as follows
\begin{eqnarray}\label{y_loyw}
(\mathbf{\Gamma}^Y_v-\sigma^2_Z \mathbf{I}_p) \mathbf{\Phi}_v = \mathbf{\gamma}^Y_v,
\end{eqnarray}
and similarly Eq. (\ref{x_loyw0}) as
\begin{eqnarray}\label{y_loyw0}
\gamma^Y(v,0) - \mathbf{\Phi}_v' \mathbf{\gamma}^Y_v = \sigma^2_\xi + \sigma^2_Z .
\end{eqnarray}
For high-order Yule-Walker equations, similarly, let us define $_s\tilde{\mathbf{\Gamma}}^{Y}_v$ as in Eq. (\ref{x_hoyw_def1}) and $_s\tilde{\mathbf{\gamma}}^{Y}_v$ as in Eq. (\ref{x_hoyw_def2}). Because $_s\tilde{\mathbf{\Gamma}}^{X}_v$ and $_s\tilde{\mathbf{\gamma}}^{X}_v$ do not contain any elements of the form $\gamma^X(w,0)$, we can replace each element $\gamma^X(w,k)$ with $\gamma^Y(w,k)$. Hence, we have $_s\tilde{\mathbf{\Gamma}}^{Y}_v = {}_s\tilde{\mathbf{\Gamma}}^{X}_v$ and $_s\tilde{\mathbf{\gamma}}^{Y}_v = {}_s\tilde{\mathbf{\gamma}}^{X}_v$. In other words, for each $v=1,2,\cdots,T$, for the process $\{Y_t\}$ we have the same form of high-order Yule-Walker equations as before (see Eq. (\ref{x_hoyw}))
\begin{eqnarray}\label{y_hoyw}
_s\tilde{\mathbf{\Gamma}}^{Y}_v \mathbf{\Phi}_v = {}_s\tilde{\mathbf{\gamma}}^{Y}_v.
\end{eqnarray}

\section{Autocovariance-based estimation methods for noise-corrupted PAR model}\label{methods}
In the proposed algorithms we use the following notations  
\begin{eqnarray}\label{emp_peacvf}
\hat{\gamma}^Y(w,k) = \frac{1}{N}\sum_{n=l}^r y_{nT+w} y_{nT+w-k},
\end{eqnarray}
where  $y_1,y_2,\cdots, y_{NT}$ is a zero-mean signal and
 \begin{eqnarray}\label{l1}
 l=\max\left(\left\lceil\frac{1-w}{T}\right\rceil,\left\lceil\frac{1-(w-k)}{T}\right\rceil\right), \quad  r=\min\left(\left\lfloor\frac{NT-w}{T}\right\rfloor,\left\lfloor\frac{NT-(w-k)}{T}\right\rfloor\right).
 \end{eqnarray}
 \begin{eqnarray}
 \hat{\mathbf{\Phi}}_v = [\hat{\phi}_1(v),\cdots,\hat{\phi}_p(v)]'.
\end{eqnarray}
Moreover, $\hat{\mathbf{\Gamma}}^Y_v$ is a $p\times p$ matrix of the following elements
\begin{eqnarray}\label{y_loyw_def1}
(\hat{\mathbf{\Gamma}}^Y_v)_{i,j} = \hat{\gamma}^Y(v-i,j-i),~~i,j=1,2,\cdots,p.
\end{eqnarray}
\begin{eqnarray}\label{y_loyw_def2}
\hat{\mathbf{\gamma}}^Y_v = [\hat{\gamma}^Y(v,1),\cdots,\hat{\gamma}^Y(v,p)]'
\end{eqnarray}
while ${}_s\hat{\tilde{\mathbf{\Gamma}}}^{Y}_v$ is a $s\times p$ matrix of the following elements
\begin{eqnarray}\label{y_hoyw_def1}
({}_s\hat{\tilde{\mathbf{\Gamma}}}^{Y}_v)_{i,j} = \hat{\gamma}^Y(v-j,p+i-j), ~~i=1,2,\cdots,s, ~~j = 1,2,\cdots, p.
\end{eqnarray}
\begin{eqnarray}\label{y_hoyw_def2}
_s\hat{\tilde{\mathbf{\gamma}}}^{Y}_v = [\hat{\gamma}^Y(v,p+1),\cdots,\hat{\gamma}^Y(v,p+s)]'.
\end{eqnarray}

\subsection{Method 1 - higher-order Yule-Walker algorithm}
The first proposed method (called M1) is a generalization of the approach presented in \cite{nasza_wojtek} for all orders $p$ of the considered model. It is based on high-order Yule-Walker equations for the noise-corrupted process $\{Y_t\}$. To obtain the vector of estimated coefficients $\hat{\mathbf{\Phi}}_v$ in a given season $v=1,2,\cdots,T$, one can use the system given in Eq. (\ref{y_hoyw}) for $s=p$ and replace all terms of the form $\gamma^Y(w,k)$ with their empirical counterparts $\hat{\gamma}^Y(w,k)$. 
In other words, for each $v=1,2,\cdots, T$, one should solve the following system of equations
\begin{eqnarray}\label{y_hoyw_method}
_p\hat{\tilde{\mathbf{\Gamma}}}^{Y}_v \hat{\mathbf{\Phi}}_v = {}_p\hat{\tilde{\mathbf{\gamma}}}^{Y}_v,
\end{eqnarray}
To obtain the estimators for $\sigma^2_\xi$ and $\sigma^2_Z$, one can use Eq. (\ref{y_loyw0}) and the first equation of the system given in Eq. (\ref{y_loyw})
\begin{eqnarray}\label{hoyw_sigma1}
\hat{\gamma}^Y(v,0) - \hat{\mathbf{\gamma}}^{Y\prime}_v \hat{\mathbf{\Phi}}_v = \hat{\sigma}^2_\xi(v) + \hat{\sigma}^2_Z(v),
\end{eqnarray}
\begin{eqnarray}\label{hoyw_sigma2}
\hat{\gamma}^Y(v,1) - \hat{\mathbf{\Gamma}}^Y_{v,1} \hat{\mathbf{\Phi}}_v  = -\hat{\sigma}^2_Z(v) \hat{\phi}_1(v),
\end{eqnarray}
where $\hat{\mathbf{\Gamma}}^Y_{v,1} = [\hat{\gamma}^Y(v-1,0),\cdots,\hat{\gamma}^Y(v-1,p-1)]$ and $\hat{\sigma}^2_\xi(v)$,  $\hat{\sigma}^2_Z(v)$ are estimators of $\sigma^2_\xi(v) = \mathbb{E}\xi_{nT+v}^2$ and $\sigma^2_Z(v) = \mathbb{E}Z_{nT+v}^2$, respectively. From Eq. (\ref{hoyw_sigma2}) we can easily calculate the $\hat{\sigma}^2_Z(v)$
\begin{eqnarray}\label{hoyw_sigmaZ}
\hat{\sigma}^2_Z(v) = \frac{\hat{\mathbf{\Gamma}}^Y_{v,1} \hat{\mathbf{\Phi}}_v - \hat{\gamma}^Y(v,1)  }{\hat{\phi}_1(v)},
\end{eqnarray}
and put it into Eq. (\ref{hoyw_sigma1}) to obtain $\hat{\sigma}^2_\xi(v)$. Because in our model given in Definition \ref{def:parma1} we assume that $\sigma^2_\xi$ and $\sigma^2_Z$ are not dependent on $v$,  finally we take
\begin{eqnarray}\label{hoyw_sigma3}
\hat{\sigma}^2_\xi = \frac{1}{T}\sum_{v=1}^T \hat{\sigma}^2_\xi(v),  \quad \hat{\sigma}^2_Z = \frac{1}{T}\sum_{v=1}^T \hat{\sigma}^2_Z(v).
\end{eqnarray}
The complete algorithm for the first method is presented in Algorithm \ref{algmethod1}. 
\begin{algorithm}
  \caption{Method M1}\label{algmethod1}
  \begin{algorithmic}[1]
    \State For each $v=1,\cdots,T$:
    \begin{enumerate}[itemsep=0pt,parsep=0pt,topsep=0pt,label=\footnotesize\roman*:]
    \item Construct $_p\hat{\tilde{\mathbf{\Gamma}}}^{Y}_v$ (Eq. \eqref{y_hoyw_def1}) and ${}_p\hat{\tilde{\mathbf{\gamma}}}^{Y}_v$ (Eq. \eqref{y_hoyw_def2}).
    \item Compute $\hat{\mathbf{\Phi}}_v =  (_p\hat{\tilde{\mathbf{\Gamma}}}^{Y}_v)^{-1} {}_p\hat{\tilde{\mathbf{\gamma}}}^{Y}_v$.
    \item Compute $\hat{\sigma}^2_Z(v)$ and $\hat{\sigma}^2_\xi(v)$ (Eq. \eqref{hoyw_sigmaZ} and Eq. \eqref{hoyw_sigma1}).
    \end{enumerate}
    \State Compute $\hat{\sigma}^2_Z = 1/T\sum_{v=1}^T \hat{\sigma}^2_Z(v)$.
    \State Compute $\hat{\sigma}^2_{\xi} = 1/T\sum_{v=1}^T \hat{\sigma}^2_{\xi}(v)$.
  \end{algorithmic}
\end{algorithm}

\subsection{Method 2 - errors-in-variables method}

The second proposed approach (called later M2) is based on the method presented in \cite{diversi1} for noise-corrupted autoregressive models. Similar as approach M1, this estimation procedure will be defined independently for each season $v=1,2,\cdots,T$. Let us define the following functions of the parameter $\sigma_Z^{2*}(v)$ based on low-order Yule-Walker equations for $\{Y_t\}$
\begin{eqnarray}\label{diversi1_phi}
\mathbf{\Phi}^*_v(\sigma^{2*}_Z(v)) = (\mathbf{\Gamma}^Y_v-\sigma^{2*}_Z(v) \mathbf{I}_p)^{-1}  \mathbf{\gamma}^Y_v,
\end{eqnarray}
\begin{eqnarray}\label{diversi1_sigmaxi}
\sigma^{2*}_\xi(v)(\sigma^{2*}_Z(v)) = \gamma^Y(v,0) - \mathbf{\Phi}^*_v(\sigma^{2*}_Z(v))' \mathbf{\gamma}^Y_v  - \sigma^{2*}_Z(v).
\end{eqnarray}
Let us also define the following matrices
\begin{eqnarray}\label{g_mat}
\mathbf{G}_v^Y = 
\begin{bmatrix}
\gamma^Y(v,0) & \mathbf{\gamma}_v^{Y\prime} \\
\mathbf{\gamma}^Y_v & \mathbf{\Gamma}^Y_v
\end{bmatrix},\quad
\mathbf{G}_v^X = 
\begin{bmatrix}
\mathbf{\gamma}^X(v,0) & \mathbf{\gamma}_v^{X\prime} \\
\mathbf{\gamma}^X_v & \mathbf{\Gamma}^X_v
\end{bmatrix}.
\end{eqnarray}
Let us note that $\mathbf{G}_v^Y = \mathbf{G}_v^X + \sigma^2_Z \mathbf{I}_{p+1}$. Hence,
\begin{eqnarray}\label{mineig}
\min \text{eig}(\mathbf{G}^Y_v) = \min \text{eig}(\mathbf{G}^X_v) + \sigma^2_Z.
\end{eqnarray}
Because both $\mathbf{G}^Y_v$ and $\mathbf{G}^X_v$ are symmetric and positive-definite (as they are autocovariance matrices of random vectors $[Y_{nT+v},\cdots,Y_{nT+v-p}]$ and $[X_{nT+v},\cdots,X_{nT+v-p}]$, respectively), we consider the values of $\sigma^{2*}_Z(v)$ from range $[0, \min \text{eig}(\mathbf{G}^Y_v)]$. In practice, we use the empirical version of $\mathbf{G}^Y_v$ defined as
\begin{eqnarray}\label{gY}
\hat{\mathbf{G}}_v^Y = 
\begin{bmatrix}
\hat{\gamma}^Y(v,0) & \hat{\mathbf{\gamma}}_v^{Y\prime} \\
\hat{\mathbf{\gamma}}^Y_v & \hat{\mathbf{\Gamma}}^Y_v
\end{bmatrix}.
\end{eqnarray}
One can see that for the true value of additive noise variance $\sigma^{2*}_Z(v) = \sigma^2_Z(v)$ we obtain $\mathbf{\Phi}^*_v(\sigma^{2}_Z(v)) = \mathbf{\Phi}_v$ and $\sigma^{2*}_\xi(v)(\sigma^{2}_Z(v)) = \sigma^{2}_\xi(v)$. Hence, given the estimate $\hat{\sigma}^{2}_Z(v)$, one can calculate all other desired estimators by utilizing it in low-order Yule-Walker equations for a noise-corrupted time series. The estimate $\hat{\sigma}^{2}_Z(v)$ is found using $s$ high-order Yule-Walker equations (where $s\geqslant p$). Namely, it is the value which minimizes the following cost function
\begin{eqnarray}\label{costhoyw}
 J_v({\sigma}^{2*}_Z(v)) = || {}_s\hat{\tilde{\mathbf{\Gamma}}}^{Y}_v \hat{\mathbf{\Phi}}^*_v({\sigma}^{2*}_Z(v)) - {}_s\hat{\tilde{\mathbf{\gamma}}}^{Y}_v ||^2_2,
\end{eqnarray}
where ${\sigma}^{2*}_Z(v) \in [0, \min \text{eig}(\hat{\mathbf{G}}^Y_v)]$ and
\begin{eqnarray}
\hat{\mathbf{\Phi}}^*_v({\sigma}^{2*}_Z(v)) = (\hat{\mathbf{\Gamma}}^Y_v-\sigma^{2*}_Z(v) \mathbf{I}_p)^{-1}  \hat{\mathbf{\gamma}}^Y_v.
\end{eqnarray}
After calculation of $\hat{\sigma}^{2}_Z(v)$, we put
\begin{eqnarray}
\hat{\mathbf{\Phi}}_v = (\hat{\mathbf{\Gamma}}^Y_v-\hat{\sigma}^{2}_Z(v) \mathbf{I}_p)^{-1}  \hat{\mathbf{\gamma}}^Y_v,
\end{eqnarray}
\begin{eqnarray}
\hat{\sigma}^{2}_\xi(v) = \hat{\gamma}^Y(v,0) - \hat{\mathbf{\Phi}}_v' \hat{\mathbf{\gamma}}^Y_v - \hat{\sigma}^{2}_Z(v).
\end{eqnarray}
At the end, as mentioned before, we can calculate $\hat{\sigma}^2_\xi$ and $\hat{\sigma}^2_Z$ as means of values obtained in each season $v$, see Eq. (\ref{hoyw_sigma3}). The details of the algorithm for M2 approach are presented in Algorithm \ref{algmethod2}. 
\begin{algorithm}
  \caption{Method M2}\label{algmethod2}
  \begin{algorithmic}[1]
    \State Set value of $s$ (where $s\geqslant p$).
    \State For each $v=1,\cdots,T$:
    \begin{enumerate}[itemsep=0pt,parsep=0pt,topsep=0pt,label=\footnotesize\roman*:]
    \item Construct $\hat{\mathbf{\Gamma}}^Y_v$ (Eq. \eqref{y_loyw_def1}), $\hat{\mathbf{\gamma}}^Y_v$ (Eq. \eqref{y_loyw_def2}), ${}_s\hat{\tilde{\mathbf{\Gamma}}}^{Y}_v$ (Eq. \eqref{y_hoyw_def1}) and ${}_s\hat{\tilde{\mathbf{\gamma}}}^{Y}_v$ (Eq. \eqref{y_hoyw_def2}).
    \item Construct $\hat{\mathbf{G}}^Y_v$ (Eq. \eqref{gY}) and compute $\min \text{eig}(\hat{\mathbf{G}}^Y_v)$.
    \item Determine $\hat{\sigma}^{2}_Z(v)$ -- a value which minimizes $J_v({\sigma}^{2*}_Z(v))$ (Eq. \eqref{costhoyw}) over interval ${\sigma}^{2*}_Z(v) \in [0, \min \text{eig}(\hat{\mathbf{G}}^Y_v)]$.
    \item Compute $\hat{\mathbf{\Phi}}_v = (\hat{\mathbf{\Gamma}}^Y_v-\hat{\sigma}^{2}_Z(v) \mathbf{I}_p)^{-1}  \hat{\mathbf{\gamma}}^Y_v$.
    \item Compute $\hat{\sigma}^{2}_\xi(v) = \hat{\gamma}^Y(v,0) - \hat{\mathbf{\Phi}}_v' \hat{\mathbf{\gamma}}^Y_v - \hat{\sigma}^{2}_Z(v)$.
    \end{enumerate}
    \State Compute $\hat{\sigma}^2_Z = 1/T\sum_{v=1}^T \hat{\sigma}^2_Z(v)$.
    \State Compute $\hat{\sigma}^2_{\xi} = 1/T\sum_{v=1}^T \hat{\sigma}^2_{\xi}(v)$.
  \end{algorithmic}
\end{algorithm}

\subsection{Method 3 - modified errors-in-variables method }

The next proposed approach (called M3) is a modification of the algorithm M2. Let us recall that the estimation procedure in Method 2 is performed completely separately for each season $v=1,2,\cdots,T$. In particular, for each $v$ we find different value of $\hat{\sigma}^2_Z(v)$ which is next utilized in calculations of other estimators. However, due to the assumption that the additive noise variance $\sigma^2_Z$ is independent of $v$, one can consider a slightly different approach. Instead of looking for $\hat{\sigma}^2_Z(v)$ in each season separately, one can calculate one value $\hat{\sigma}^2_Z$ for all seasons at once and then use it for each $v$ in the next stage. Now, this estimate is a value which minimizes the following total cost
\begin{eqnarray}\label{jtotal}
J_{\text{total}}(\sigma^{2*}_Z) = \sum_{v=1}^T J_v(\sigma^{2*}_Z),
\end{eqnarray}
where $J_v(\sigma^{2*}_Z)$ has the same form as in M2 approach, see Eq. (\ref{costhoyw}). We consider $\sigma^{2*}_Z \in [0, \zeta]$, where $\zeta$ is the minimum of values $\min \text{eig}(\hat{\mathbf{G}}^Y_v)$ for $v=1,2,\cdots,T$. With $\hat{\sigma}^2_Z$ obtained, one can calculate $\hat{\mathbf{\Phi}}_v$ and $\hat{\sigma}^{2}_\xi(v)$ using the following
\begin{eqnarray}
\hat{\mathbf{\Phi}}_v = (\hat{\mathbf{\Gamma}}^Y_v-\hat{\sigma}^{2}_Z \mathbf{I}_p)^{-1}  \hat{\mathbf{\gamma}}^Y_v,
\end{eqnarray}
\begin{eqnarray}
\hat{\sigma}^{2}_\xi(v) = \hat{\gamma}^Y(v,0) - \hat{\mathbf{\Phi}}_v' \hat{\mathbf{\gamma}}^Y_v - \hat{\sigma}^{2}_Z,
\end{eqnarray}
for each $v=1,2,\cdots,T$. Similarly as previously, we can calculate $\hat{\sigma}_\xi^2$ as mean of $\hat{\sigma}_\xi^2(v)$ values. The detailed algorithm of the approach M3 we present in Algorithm \ref{algmethod3}. 
\begin{algorithm}
  \caption{Method M3}\label{algmethod3}
  \begin{algorithmic}[1]
    \State Set value of $s$ (where $s\geqslant p$).
    \State For each $v=1,\cdots,T$:
    \begin{enumerate}[itemsep=0pt,parsep=0pt,topsep=0pt,label=\footnotesize\roman*:]
    \item Construct $\hat{\mathbf{\Gamma}}^Y_v$ (Eq. \eqref{y_loyw_def1}), $\hat{\mathbf{\gamma}}^Y_v$ (Eq. \eqref{y_loyw_def2}), ${}_s\hat{\tilde{\mathbf{\Gamma}}}^{Y}_v$ (Eq. \eqref{y_hoyw_def1}) and ${}_s\hat{\tilde{\mathbf{\gamma}}}^{Y}_v$ (Eq. \eqref{y_hoyw_def2}).
    \item Construct $\hat{\mathbf{G}}^Y_v$ (Eq. \eqref{gY}) and compute $\min \text{eig}(\hat{\mathbf{G}}^Y_v)$.
    \end{enumerate}
    \State Compute $\zeta = \min \left\{ \min \text{eig}(\hat{\mathbf{G}}^Y_v),\: :\: v=1,\cdots,T \right\}$
    \State Determine $\hat{\sigma}^2_Z$ -- a value which minimizes $J_{\text{total}}(\sigma^{2*}_Z)$ (Eq. \eqref{jtotal}) over interval $\sigma^{2*}_Z \in [0, \zeta]$.
    \State For each $v=1,\cdots,T$ :
        \begin{enumerate}[itemsep=0pt,parsep=0pt,topsep=0pt,label=\footnotesize\roman*:]
        \item Compute $\hat{\mathbf{\Phi}}_v = (\hat{\mathbf{\Gamma}}^Y_v-\hat{\sigma}^{2}_Z \mathbf{I}_p)^{-1}  \hat{\mathbf{\gamma}}^Y_v$.
        \item Compute $\hat{\sigma}^{2}_\xi(v) = \hat{\gamma}^Y(v,0) - \hat{\mathbf{\Phi}}_v' \hat{\mathbf{\gamma}}^Y_v - \hat{\sigma}^{2}_Z$.
        \end{enumerate}
    \State Compute $\hat{\sigma}^2_{\xi} = 1/T\sum_{v=1}^T \hat{\sigma}^2_{\xi}(v)$
  \end{algorithmic}
\end{algorithm}

\subsection{Method 4 - constrained least squares optimization-based method}

The last approach proposed in this paper (called later M4) is based on the method introduced in \cite{esfandiari} for autoregressive processes with additive noise. In this case, the procedure is again performed independently for each season $v=1,2,\cdots,T$. Here, the first step is to solve an optimization problem, where the least squares-type cost is constructed using low-order Yule-Walker equations and the constraint is equal to the first high-order Yule-Walker equation. In our case, for noise-corrupted periodic autoregressive processes, this problem is defined in the following way
\begin{eqnarray}\label{esf_problem}
\min_{\hat{\mathbf{\Phi}}_v,\,\hat{\sigma}^2_Z(v)} ||(\hat{\mathbf{\Gamma}}^Y_v-\hat{\sigma}^2_Z(v) \mathbf{I}_p) \hat{\mathbf{\Phi}}_v - \hat{\mathbf{\gamma}}^Y_v||^2_2 \qquad
\text{subject to} \quad  \hat{\tilde{\mathbf{\Gamma}}}^{Y}_{v,1} \hat{\mathbf{\Phi}}_v = \hat{\gamma}^Y(v,p+1),
\end{eqnarray}
where $\hat{\tilde{\mathbf{\Gamma}}}^Y_{v,1} = [\hat{\gamma}^Y(v-1,p),\hat{\gamma}^Y(v-2,p-1),\cdots,\hat{\gamma}^Y(v-p,1)]$. To solve the stated optimization problem, one can use the method presented in \cite{esfandiari}. It is an iterative method, where we alternately calculate values of $\hat{\sigma}_Z^2(v)$ and $\hat{\mathbf{\Phi}}_v$ using the following formulas
\begin{eqnarray}\label{iter_formula_sig}
\hat{\sigma}_Z^2(v)^{(i)} = \frac{\hat{\mathbf{\Phi}}_v^{(i-1)\prime}\left(\hat{\mathbf{\Gamma}}^Y_v\hat{\mathbf{\Phi}}_v^{(i-1)} - \hat{\mathbf{\gamma}}^Y_v\right)}{\left|\left|\hat{\mathbf{\Phi}}_v^{(i-1)}\right|\right|^2_2}
\end{eqnarray}
\begin{eqnarray}\label{iter_formula_phi}
\hat{\mathbf{\Phi}}_v^{(i)} = \left(\hat{\mathbf{\Gamma}}^Y_v -\hat{\sigma}_Z^2(v)^{(i)} \mathbf{I}_p\right)^{-1}\hat{\mathbf{\gamma}}^Y_v - \left(\frac{\hat{\tilde{\mathbf{\Gamma}}}^{Y}_{v,1} \left(\hat{\mathbf{\Gamma}}^Y_v -\hat{\sigma}_Z^2(v)^{(i)} \mathbf{I}_p\right)^{-1}\hat{\mathbf{\gamma}}^Y_v-\hat{\gamma}^Y(v,p+1)}{\hat{\tilde{\mathbf{\Gamma}}}^{Y}_{v,1}\left[\left(\hat{\mathbf{\Gamma}}^Y_v -\hat{\sigma}_Z^2(v)^{(i)} \mathbf{I}_p\right)^2\right]^{-1}\hat{\tilde{\mathbf{\Gamma}}}^{Y\prime}_{v,1}}\right)\left[\left(\hat{\mathbf{\Gamma}}^Y_v -\hat{\sigma}_Z^2(v)^{(i)} \mathbf{I}_p\right)^2\right]^{-1}\hat{\tilde{\mathbf{\Gamma}}}^{Y\prime}_{v,1}
\end{eqnarray}
until a convergence criterion is fulfilled. After this procedure, we use the obtained $\hat{\sigma}_Z^2(v)$ in the estimation of $\mathbf{\Phi}_v$. Here, it is done using both low- and high-order Yule-Walker equations. We consider the following system of equations
\begin{eqnarray}\label{h_system}
\hat{\mathbf{H}}_v \hat{\mathbf{\Phi}}_v = \hat{\mathbf{h}}_v
\end{eqnarray}
where
\begin{eqnarray}\label{M4_Hvecmat}
    \hat{\mathbf{H}}_v = \begin{bmatrix}
\hat{\mathbf{\Gamma}}^Y_v-\hat{\sigma}^{2}_Z(v) \mathbf{I}_p\\
{}_s\hat{\tilde{\mathbf{\Gamma}}}^{Y}_v
\end{bmatrix}, \quad
     \hat{\mathbf{h}}_v = \begin{bmatrix}
    \hat{\mathbf{\gamma}}^Y_v\\
    {}_s\hat{\tilde{\mathbf{\gamma}}}^{Y}_v
    \end{bmatrix}.
\end{eqnarray}
To obtain  $\hat{\mathbf{\Phi}}_v$, we calculate the least-squares solution of system given in Eq. \eqref{h_system}
\begin{eqnarray}\label{lssolution}
\hat{\mathbf{\Phi}}_v = (\hat{\mathbf{H}}_v'\hat{\mathbf{H}}_v)^{-1}\hat{\mathbf{H}}_v'\hat{\mathbf{h}}_v.
\end{eqnarray}
We also estimate $\hat{\sigma}_{\xi}^2(v)$ using the same equation as before
\begin{eqnarray}
\hat{\sigma}^{2}_\xi(v) = \hat{\gamma}^Y(v,0) - \hat{\mathbf{\Phi}}_v' \hat{\mathbf{\gamma}}^Y_v - \hat{\sigma}^{2}_Z(v).
\end{eqnarray}
The detailed algorithm for the method M4 is presented in Algorithm \ref{algmethod4}. Here, we also utilize the algorithm for finding the initial value of $\sigma^2_Z(v)$ presented in \cite{esfandiari}. {In point 2.ii.a of our algorithm, one can use some other number closer to 1 instead of 0.9999, to achieve that $f(D_1)$ and $f(D_2)$ would have different signs.}

\begin{algorithm}
  \caption{Method M4}\label{algmethod4}
  \begin{algorithmic}[1]
    \State Set values of $s$, $\delta_0$ and $\delta$ (small positive numbers).
    \State For each $v=1,\cdots,T$:
        \begin{enumerate}[itemsep=0pt,parsep=0pt,topsep=0pt,label=\footnotesize\roman*:]
    \item Construct $\hat{\mathbf{\Gamma}}^Y_v$ (Eq. \eqref{y_loyw_def1}), $\hat{\mathbf{\gamma}}^Y_v$ (Eq. \eqref{y_loyw_def2}), ${}_s\hat{\tilde{\mathbf{\Gamma}}}^{Y}_v$ (Eq. \eqref{y_hoyw_def1}), ${}_s\hat{\tilde{\mathbf{\gamma}}}^{Y}_v$ (Eq. \eqref{y_hoyw_def2}) and $\tilde{\mathbf{\Gamma}}^Y_{v,1}$ (Eq. \eqref{esf_problem}).
    \item Find $\hat{\sigma}_Z^2(v)^{(0)}$, initial value of $\hat{\sigma}^2_Z(v)$:
            \begin{enumerate}[itemsep=0pt,parsep=0pt,topsep=0pt,label=\footnotesize\alph*:]
            \item Set $D_1=0$ and $D_2=0.9999 \min \text{eig}(\hat{\mathbf{\Gamma}}^Y_v)$.
            \item Compute $D=(D_1+D_2)/2$ and $f(D) = \hat{\gamma}^Y(v,0) - D - \hat{\mathbf{\gamma}}_v^{Y\prime}(\hat{\mathbf{\Gamma}}_v^Y-D \mathbf{I}_p)^{-1}\hat{\mathbf{\gamma}}_v^{Y}$. If $|f(D)|\leqslant \delta_0$, set $\hat{\sigma}_Z^2(v)^{(0)} = D$ and go to point 2.iii.
            \item If $f(D)>0$, set $D_1=D$, and if $f(D)<0$, set $D_2=D$ and go to point 2.ii.b.
            \end{enumerate}
    \item Set $i=0$. 
    \item Compute $\hat{\mathbf{\Phi}}_v^{(i)}$ (Eq. \eqref{iter_formula_phi}).
    \item Set $i = i+1$ and compute $\hat{\sigma}^2_Z(v)^{(i)}$ (Eq. \eqref{iter_formula_sig}). If $\frac{\left|\hat{\sigma}^2_Z(v)^{(i)} - \hat{\sigma}^2_Z(v)^{(i-1)} \right|}{\left|\hat{\sigma}^2_Z(v)^{(i-1)}\right|}\leqslant \delta$, go to point 2.vi. Otherwise, go to point 2.iv.
    \item Compute $\hat{\mathbf{H}}_v$ and $\hat{\mathbf{h}}_v$ (Eq. \eqref{M4_Hvecmat}), where $\hat{\sigma}_Z^2(v) = \hat{\sigma}_Z^2(v)^{(i)}$.
    \item Compute $\hat{\mathbf{\Phi}}_v = (\hat{\mathbf{H}}_v'\hat{\mathbf{H}}_v)^{-1}\hat{\mathbf{H}}_v'\hat{\mathbf{h}}_v$.
    \item Compute $\hat{\sigma}^{2}_\xi(v) = \hat{\gamma}^Y(v,0) - \hat{\mathbf{\Phi}}_v' \hat{\mathbf{\gamma}}^Y_v - \hat{\sigma}^{2}_Z(v)$.
        \end{enumerate}
    \State Compute $\hat{\sigma}^2_Z = 1/T\sum_{v=1}^T \hat{\sigma}^2_Z(v)$.
    \State Compute $\hat{\sigma}^2_{\xi} = 1/T\sum_{v=1}^T \hat{\sigma}^2_{\xi}(v)$
  \end{algorithmic}
\end{algorithm}

\section{Analysis of the simulated data - PAR model with Gaussian additive noise}\label{simul1}

\begin{figure}
\centering
    \includegraphics[width=\textwidth]{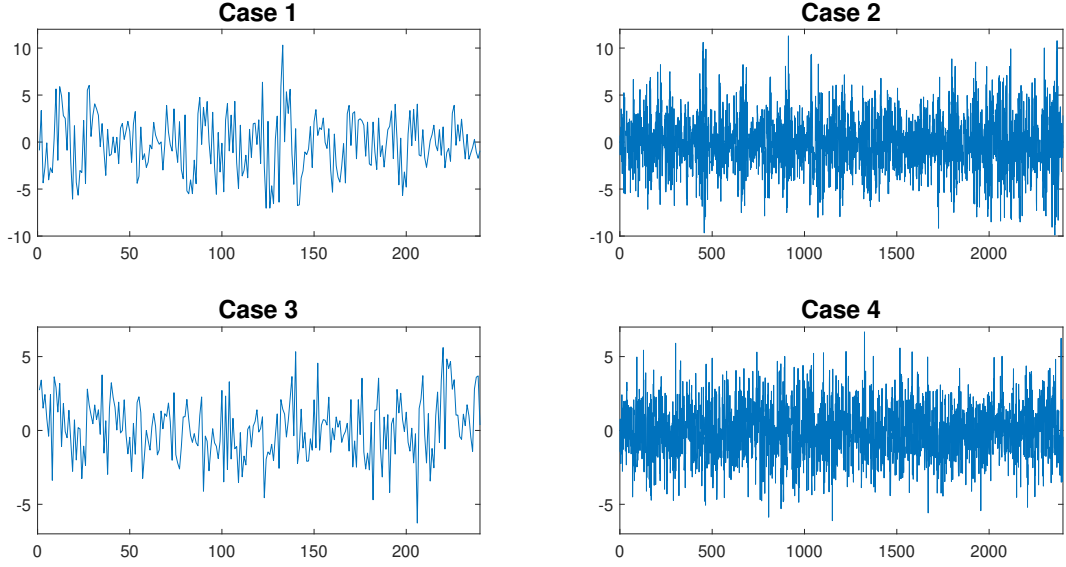}
      \caption{Sample trajectories of PAR model with Gaussian additive noise related to the cases considered in the simulation study.}
      \label{traj1}
\end{figure}

In this section, we check the performance of the methods introduced in Section \ref{methods} on artificial data using Monte Carlo simulations. We assume that $T$ and $p$ are known and we focus on the estimation of $\phi_i(v)$ coefficients for $i=1,\cdots,p$ and $v=1,\cdots,T$. For the comparison, we also analyze the classical Yule-Walker method (M5) as an approach that does not take into account the assumption of additive noise presence. This method is based on the low-order Yule-Walker equations for the pure PAR process $\{X_t\}$ or, equivalently, for $\{Y_t\}$ with $\sigma^2_Z = 0$ defined in Eq. (\ref{model1}). In the classical Yule-Walker method for each $v=1,2,\cdots, T$, to obtain $\hat{\mathbf{\Phi}}_v$, one should solve the following system of equations
\begin{eqnarray}\label{classicalyw}
\hat{\mathbf{\Gamma}}^Y_v \hat{\mathbf{\Phi}}_v = \hat{\mathbf{\gamma}}^Y_v.
\end{eqnarray}
The general set-up for the simulations is as follows. We consider the randomly generated trajectories from the PAR model with Gaussian additive noise with  variance $\sigma^2_Z=0.8$. The innovations of the underlying PAR model $\{\xi_t\}$ have a Gaussian distribution with variance $\sigma^2_\xi=1$. We consider the model with $p=2$, $T=3$ and the following fixed values of coefficients: $\phi_1(1)=0.6$, $\phi_1(2)=-0.9$, $\phi_1(3)=-0.5$, $\phi_2(2)=1.4$, $\phi_2(3)=0.7$. For  coefficient $\phi_2(1)$ (not mentioned above), we consider two cases:  $\phi_2(1)=-0.8$ and $\phi_2(1)=-0.1$. The choice of such two values, where one is more distant and the second is closer to zero, is made to illustrate the drawback of the method M1 based on the high-order Yule-Walker equations in the latter case, see \cite{nasza_wojtek}, and assess the performance of other developed methods in such situation. We also analyze two lengths of a single sample trajectory, namely $NT = 240$ and $NT=2400$. Hence, in total we consider four scenarios (due to $\phi_2(1)$ and $NT$ values), namely
\begin{multicols}{2}
    \begin{itemize}
        \item Case 1: $\phi_2(1)=-0.8$, $NT=240$,
        \item Case 2: $\phi_2(1)=-0.8$, $NT=2400$,
        \item Case 3: $\phi_2(1)=-0.1$, $NT=240$,
        \item Case 4: $\phi_2(1)=-0.1$, $NT=2400$.
    \end{itemize}
\end{multicols}
As examples, single sample trajectories for each of these cases are presented in Fig. \ref{traj1}.
For each of the cases, we simulate $M=1000$ trajectories of the given set-up. For each generated sample, we estimate the $\phi_i(v)$ parameters using all five considered methods (M1-M5). In such a way, we are able to obtain many realizations of all estimators which allow us to analyze their empirical distributions. As results, for each method and each estimated coefficient, we create a boxplot of obtained values and calculate the mean squared error (MSE). As some of the presented methods have hyperparameters needed to be preset, let us mention their values applied here. For methods M2, M3 and M4, we take here $s=p=2$ high-order Yule-Walker equations. Moreover, for the method M4 we set $\delta_0 = 0.001$ and $\delta = 0.001$.

\begin{figure}
\centering
    \includegraphics[width=\textwidth]{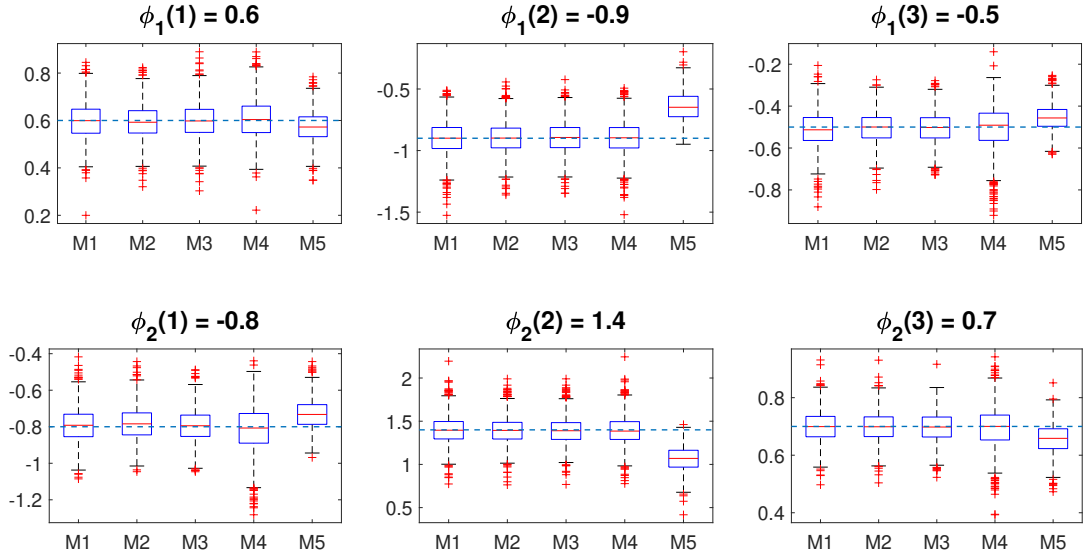}
      \caption{Boxplots of estimated values in Case 1 ($\phi_2(1) = -0.8$, $NT = 240$) for all considered methods.}
      \label{bp1}
\end{figure}

The boxplots of the estimated values for all considered methods in Case 1 are presented in Fig. \ref{bp1}. Most of all, one can see the difference between the results for the proposed methods (M1-M4) and the classical Yule-Walker method (M5) in terms of the present bias -- unlike the former, the latter seems to be significantly biased. In general, the level of this bias increases, the further the true value of the parameter is from zero. The MSE values for all methods for Case 1 are presented in Tab. \ref{tab1}. One can see the clear advantage of methods where the additive noise presence is taken into account. In particular, the best results were obtained by method M3. 

\begin{table}
\centering
\begin{tabular}{| c | c | c | c | c | c | c |c|}
\hline
method & $\phi_1(1)=0.6$ & $\phi_1(2)=-0.9$ & $\phi_1(3)=-0.5$ & $\phi_2(1)=-0.8$ & $\phi_2(2)=1.4$ & $\phi_2(3)=0.7$ & average\\
\hline

M1 & 0.0063 & 0.0188  & 0.0074 & 0.0091 & 0.0272 & 0.0030 & 0.0120 \\ \hline
M2 & 0.0056  & 0.0169  & 0.0055 & 0.0091 & \textbf{0.0258} & 0.0029 & 0.0110 \\ \hline
M3 & 0.0059  & \textbf{0.0168}  & \textbf{0.0053} & \textbf{0.0081} & \textbf{0.0258} & \textbf{0.0027} & \textbf{0.0107} \\ \hline
M4 & 0.0071 & 0.0184  & 0.0112 & 0.0158 & 0.0280 & 0.0053 & 0.0143 \\ \hline
M5 & \textbf{0.0050} & 0.0795  &0.0058  &0.0114  & 0.1348 & 0.0046 & 0.0402 \\ \hline

\end{tabular}
\caption{Parameter-wise and average mean squared errors obtained in Case 1 ($\phi_2(1)=-0.8$, $NT=240$) for all considered methods.}
\label{tab1}
\end{table}

\begin{figure}
\centering
    \includegraphics[width=\textwidth]{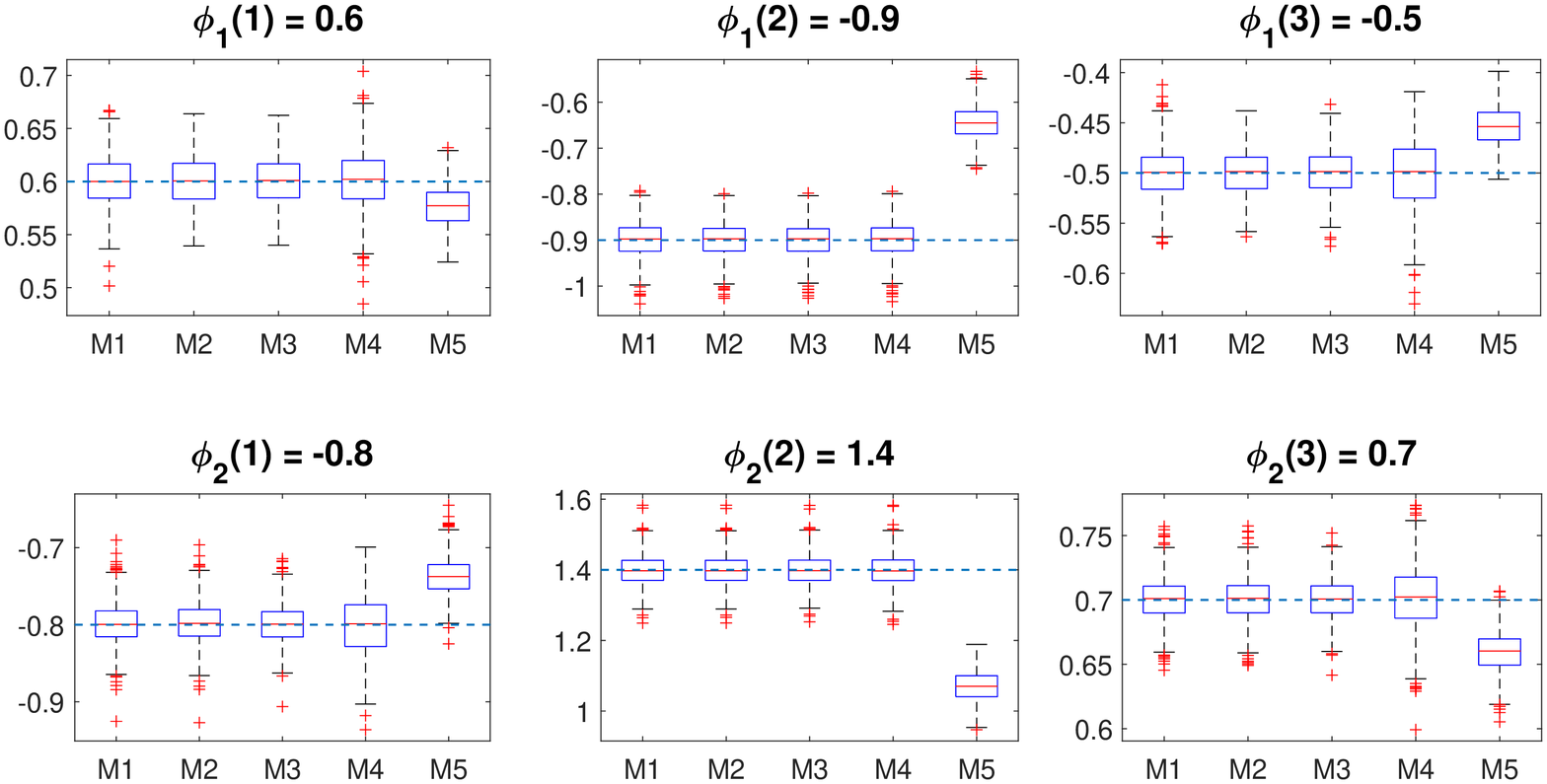}
      \caption{Boxplots of estimated values in Case 2 ($\phi_2(1) = -0.8$, $NT = 2400$) for all considered methods.}
      \label{bp2}
\end{figure}

The boxplots for the Case 2 presented in Fig. \ref{bp2} show that the aforementioned bias for the classical Yule-Walker method is still present, even though the samples under consideration are much longer. As expected, the variance of all estimators is lower than before. This can be also seen in the MSE results for this case presented in Tab. \ref{tab2}. The performance of all methods improved in comparison to Case 1, but for methods M1-M4 this change is much more significant because of a lack of bias. Similar as for Case 1, method M3 yields the best results.

\begin{table}
\centering
\begin{tabular}{| c | c | c | c | c | c | c |c|}
\hline
method & $\phi_1(1)=0.6$ & $\phi_1(2)=-0.9$ & $\phi_1(3)=-0.5$ & $\phi_2(1)=-0.8$ & $\phi_2(2)=1.4$ & $\phi_2(3)=0.7$ & average\\
\hline

M1 & 0.0006  & 0.0014 & 0.0006 & 0.0007  & \textbf{0.0019} & 0.0003 & 0.0009  \\ \hline
M2 & \textbf{0.0005}  &0.0014  &\textbf{0.0005} & 0.0007 & \textbf{0.0019} &0.0003  & 0.0009 \\ \hline
M3 & \textbf{0.0005} &\textbf{0.0013}  & \textbf{0.0005} & \textbf{0.0006} & \textbf{0.0019} & \textbf{0.0002} & \textbf{0.0008} \\ \hline
M4 & 0.0008 & 0.0014  &0.0012 &0.0015  & \textbf{0.0019} & 0.0006 & 0.0012 \\ \hline
M5 & 0.0009 & 0.0667 & 0.0025 & 0.0045  & 0.1108 & 0.0019 & 0.0312 \\ \hline

\end{tabular}
\caption{Parameter-wise and average mean squared errors obtained in Case 2 ($\phi_2(1)=-0.8$, $NT=2400$) for all considered methods.}
\label{tab2}
\end{table}

\begin{figure}
\centering
    \includegraphics[width=\textwidth]{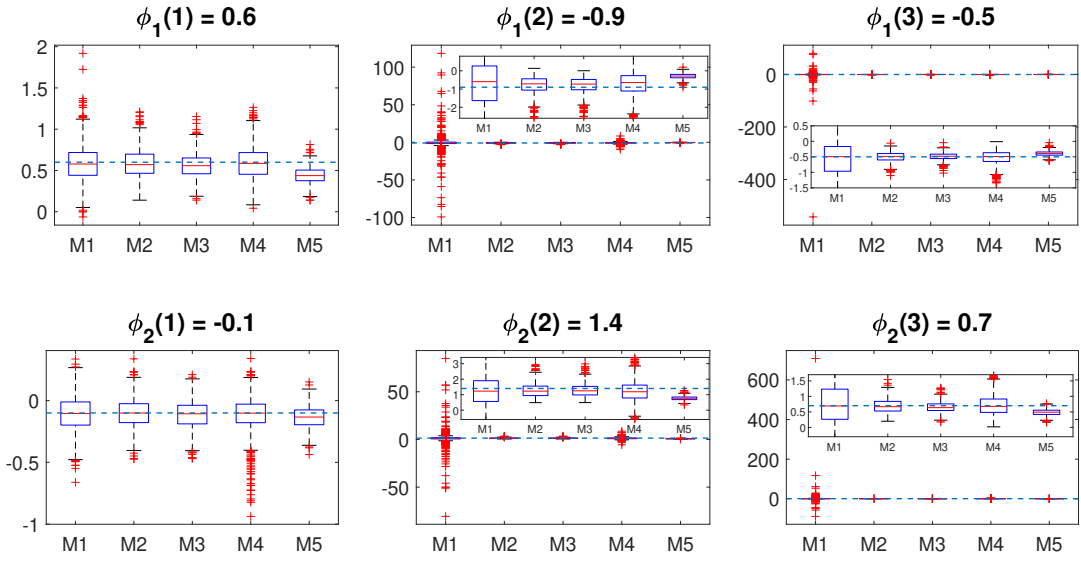}
      \caption{Boxplots of estimated values in Case 3 ($\phi_2(1) = -0.1$, $NT = 240$) for all considered methods.}
      \label{bp3}
\end{figure}

The results for Case 3 are illustrated in Fig. \ref{bp3} and presented in Tab. \ref{tab3}. Let us remind that, in comparison to previous cases, we changed the value of $\phi_2(1)$ coefficient from -0.8 to -0.1, making it much closer to zero. Above all, let us note that method M1 clearly failed here, producing very significant errors for some parameters. This behavior is caused by the fact that when the true coefficients values are close to zero, the matrix $_p\tilde{\mathbf{\Gamma}}^{Y}_v$ used in estimation with method M1 becomes close to singular. This issue is also discussed in \cite{kay1980}, in the context of noise-corrupted autoregressive models. 

\begin{figure}
\centering
    \includegraphics[width=\textwidth]{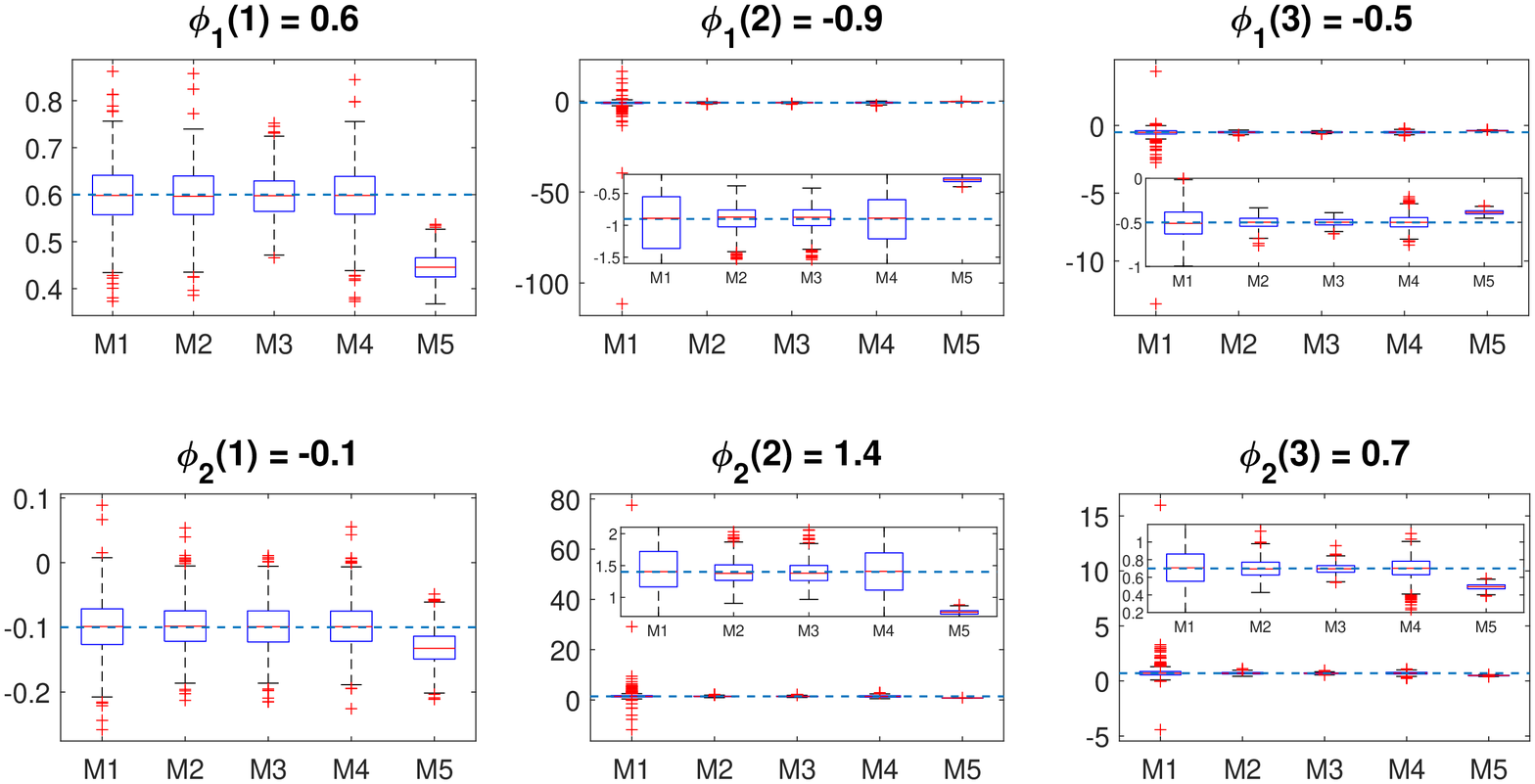}
      \caption{Boxplots of estimated values in Case 4 ($\phi_2(1) = -0.1$, $NT = 2400$) for all considered methods.}
      \label{bp4}
\end{figure}

Hence, especially for a small amount of data, this approach might fail. In Case 4, where we consider longer samples, this drawback is slightly mitigated as can be seen in Fig. \ref{bp4} and Tab. \ref{tab4}, but the method is still much less reliable than others. {Another method with visible performance drop is method M4.}{} In Case 3, its results are even worse than for method M5, but because of the previously mentioned improvement tendencies for longer samples, in Case 4 the method M4 gives again better results. However, in both setups, similarly as before, the best results are obtained by methods M2 and M3, in particular for the latter which can be clearly considered as the best approach in this simulation study.

\begin{table}
\centering
\begin{tabular}{| c | c | c | c | c | c | c |c|}
\hline
method & $\phi_1(1)=0.6$ & $\phi_1(2)=-0.9$ & $\phi_1(3)=-0.5$ & $\phi_2(1)=-0.1$ & $\phi_2(2)=1.4$ & $\phi_2(3)=0.7$ & average\\
\hline

M1 & 0.0515 & 109.97 &333.44  & 0.0210  & 51.2423 & 546.18 & 173.49  \\ \hline
M2 & 0.0296  &0.2278 & 0.0225 & 0.0139  & 0.1972 & 0.0466 & 0.0896  \\ \hline
M3 & \textbf{0.0216}  &\textbf{0.2008} & \textbf{0.0135} & 0.0125  &\textbf{0.1781}  & \textbf{0.0270} & \textbf{0.0756} \\ \hline
M4 & 0.0395 &1.1913 & 0.0713 & 0.0215  & 0.8316 & 0.1365 & 0.3819  \\ \hline
M5 &  0.0359 &0.3895 &0.0181  &\textbf{0.0090}   & 0.4159 & 0.0557 & 0.1540  \\ \hline

\end{tabular}
\caption{Parameter-wise and average mean squared errors obtained in Case 3 ($\phi_2(1)=-0.1$, $NT=240$) for all considered methods.}
\label{tab3}
\end{table}

\begin{table}
\centering
\begin{tabular}{| c | c | c | c | c | c | c |c|}
\hline
method & $\phi_1(1)=0.6$ & $\phi_1(2)=-0.9$ & $\phi_1(3)=-0.5$ & $\phi_2(1)=-0.1$ & $\phi_2(2)=1.4$ & $\phi_2(3)=0.7$ & average\\
\hline

M1 & 0.0042 & 15.625 & 0.2566 & 0.0019  & 7.4705 & 0.3709 & 3.9549  \\ \hline
M2 & 0.0038 & 0.0442& 0.0043 & \textbf{0.0013}  & 0.0373 & 0.0112 & 0.0170  \\ \hline
M3 & \textbf{0.0023} &\textbf{0.0363} &\textbf{0.0017}  & \textbf{0.0013}  & \textbf{0.0314} & \textbf{0.0029} & \textbf{0.0127}  \\ \hline
M4 & 0.0037 & 0.1798& 0.0066 & \textbf{0.0013}  & 0.1582 & 0.0150 & 0.0608  \\ \hline
M5 & 0.0248 & 0.3792& 0.0138 & 0.0017  & 0.4056 & 0.0444 & 0.1449  \\ \hline

\end{tabular}
\caption{Parameter-wise and average mean squared errors obtained in Case 4 ($\phi_2(1)=-0.1$, $NT=2400$) for all considered methods.}
\label{tab4}
\end{table}
\section{Analysis of the simulated data - PAR model with  additive outliers}\label{sec_outl}
To demonstrate the universality of the proposed estimation techniques, in this section we analyze two additional cases of the additive noise $\{Z_t\}$ in the model (\ref{model1}), namely when  it is a sequence of additive outliers and when it is a mixture of the Gaussian additive noise and additive outliers. As it was mentioned in Section \ref{intro}, especially the last case corresponds to real situations, where the measurements are disturbed by the additive errors (related to the noise of the device) and additional impulses that may be related to external sources.   

In the presented simulation study, we consider the PAR model corresponding to Cases 1 and 2 described in Section \ref{simul1}. In order to avoid the comprehensive discussion of the influence of the level of the additive noise's variance on the estimation results, we assume the variance of the additive noise considered in this part is on the same level as the variance of the Gaussian additive noise considered in the previous section ($\sigma_Z^2=0.8$).  The analysis related to the sensitivity of the new estimation techniques to the variance of the additive noise is an essential issue and will be discussed in our future studies. In this part, we only analyze how the existence of additive outliers in the considered model influences the estimation. Similar as in Section \ref{simul1} we provide the Monte Carlo simulations and in each case we simulate $M=1000$ trajectories corresponding to the considered cases. For each simulated trajectory, we estimate the parameters of the model, finally we create the boxplots of the received estimators and calculate the  average mean squared errors.   

\begin{figure}
\centering
    \includegraphics[width=\textwidth]{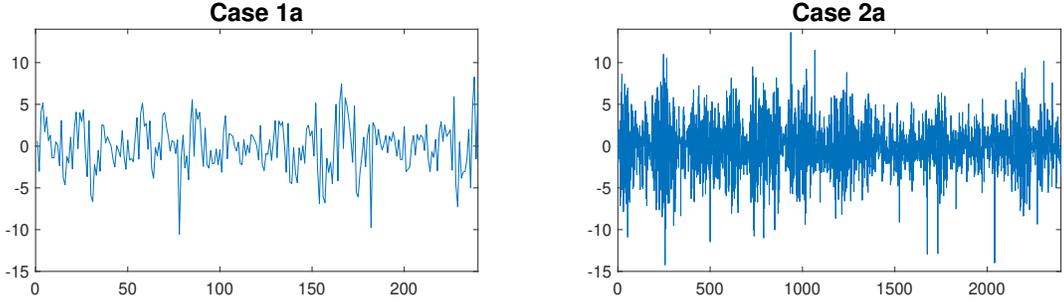}
      \caption{Sample trajectories of PAR model with additive outliers related to the Cases 1a and 2a considered in the simulation study.}
      \label{traj2}
\end{figure}

First, we analyze the case when the additive noise $\{Z_t\}$  in the model (\ref{model1}) is a sequence of additive outliers. More precisely, we assume that for each $t$, $\mathbb{P}(Z_t=10)=\mathbb{P}(Z_t=-10)=0.004$ and $\mathbb{P}(Z_t=0)=0.992$. In the further analysis the considered cases are called Case 1a and Case 2a, respectively, in order to highlight that they correspond to the Case 1 and Case 2, analyzed in the previous section.  In Fig. \ref{traj2} we demonstrate the exemplary trajectories of the considered model corresponding to Case 1a and Case 2a. 

\begin{figure}
\centering
    \includegraphics[width=\textwidth]{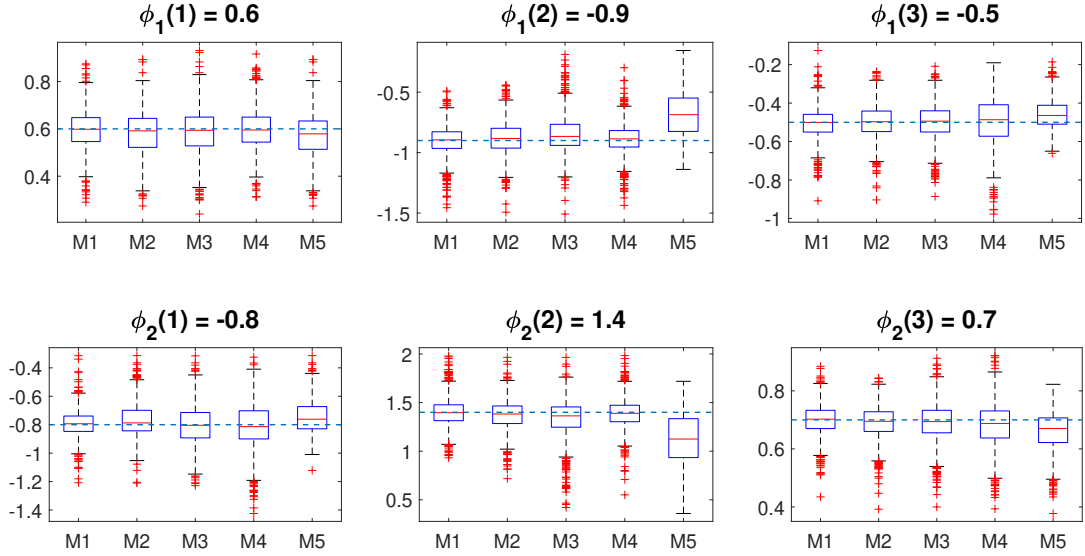}
      \caption{Boxplots of estimated values in Case 1a ($\phi_2(1) = -0.8$, $NT = 240$) for all considered methods.}
      \label{aobp1}
\end{figure}

The boxplots obtained for the Case 1a are presented in Fig. \ref{aobp1}. Most of all, they confirm that the developed methods are able to handle also the cases where the additive outliers are present. In general, one can observe the similar behavior of results as the one presented in Section \ref{simul1}. While methods M1-M4 does not seem to possess any significant bias, for the classical Yule-Walker method (M5) once again it is not a case. Hence, as shown by the mean squared errors in Tab. \ref{aotab1}, the method M5 performed much worse than others. However, one can see some changes of tendencies in comparison to the cases with Gaussian additive noise. In particular, the method M3, which yielded the best results in Section \ref{simul1}, here turned out to be less efficient than other introduced methods. The least average mean squared error was obtained for method M1.

\begin{table}
\centering
\begin{tabular}{| c | c | c | c | c | c | c |c|}
\hline
method & $\phi_1(1)=0.6$ & $\phi_1(2)=-0.9$ & $\phi_1(3)=-0.5$ & $\phi_2(1)=-0.8$ & $\phi_2(2)=1.4$ & $\phi_2(3)=0.7$ & average\\
\hline

M1 & \textbf{0.0062} & \textbf{0.0158} & \textbf{0.0064} & \textbf{0.0086}  & \textbf{0.0218} & \textbf{0.0029} & \textbf{0.0103}  \\ \hline
M2 & 0.0090 & 0.0189 & 0.0079 & 0.0153 & 0.0229 & 0.0032 & 0.0129  \\ \hline
M3 & 0.0099 & 0.0257 & 0.0081 & 0.0204 & 0.0402 & 0.0044 & 0.0181  \\ \hline
M4 & 0.0073 & 0.0161 & 0.0148 & 0.0256 & 0.0237 & 0.0058 & 0.0156  \\ \hline
M5 & 0.0091 & 0.0828 & 0.0070 & 0.0165  & 0.1406 & 0.0057 & 0.0436  \\ \hline

\end{tabular}
\caption{Parameter-wise and average mean squared errors obtained in Case 1a ($\phi_2(1)=-0.8$, $NT=240$) for all considered methods.}
\label{aotab1}
\end{table}

\begin{figure}
\centering
    \includegraphics[width=\textwidth]{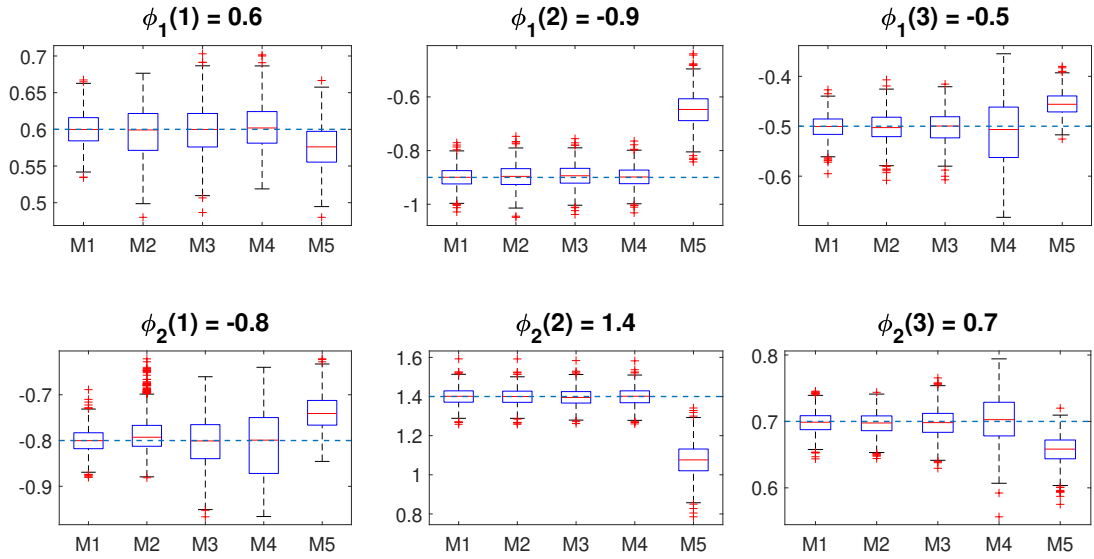}
      \caption{Boxplots of estimated values in Case 2a ($\phi_2(1) = -0.8$, $NT = 2400$) for all considered methods.}
      \label{aobp2}
\end{figure}

Next, let us analyze the Case 2a, where we consider longer samples. As can be seen in Fig. \ref{aobp2} and Tab. \ref{aotab2}, the results indicate a higher efficiency than in the previous case, but the values obtained with the method M5 still exhibit its bias towards zero. Hence, similarly as before, for longer samples the advantage of proposed methods becomes even more significant. The best results, as in Case 1a, were obtained by method M1. However, one should be aware of the fact that for lower true values of $\phi_i(v)$ coefficients (see Cases 3 and 4 in Section \ref{simul1}) this method would still suffer from the mentioned causes.
\begin{table}
\centering
\begin{tabular}{| c | c | c | c | c | c | c |c|}
\hline
method & $\phi_1(1)=0.6$ & $\phi_1(2)=-0.9$ & $\phi_1(3)=-0.5$ & $\phi_2(1)=-0.8$ & $\phi_2(2)=1.4$ & $\phi_2(3)=0.7$ & average\\
\hline

M1 & \textbf{0.0005} & \textbf{0.0013} & \textbf{0.0005} & \textbf{0.0007}  & \textbf{0.0019} & \textbf{0.0003} & \textbf{0.0009}  \\ \hline
M2 & 0.0013 & 0.0018 & 0.0009 & 0.0019  & \textbf{0.0019}  & \textbf{0.0003} & 0.0013  \\ \hline
M3 & 0.0011 & 0.0016 & 0.0010  & 0.0028  & 0.0020  & 0.0005 & 0.0015  \\ \hline
M4 & 0.0010& \textbf{0.0013} & 0.0040  & 0.0052  & 0.0020  & 0.0012 & 0.0024  \\ \hline
M5 & 0.0015 & 0.0678 & 0.0025  &0.0051  &0.1127  & 0.0023 & 0.0320  \\ \hline

\end{tabular}
\caption{Parameter-wise and average mean squared errors obtained in Case 2a ($\phi_2(1)=-0.8$, $NT=2400$) for all considered methods.}
\label{aotab2}
\end{table}

\begin{figure}
\centering
    \includegraphics[width=\textwidth]{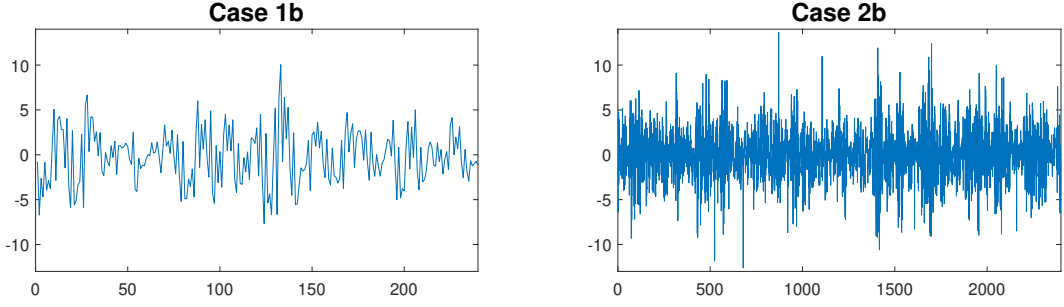}
      \caption{Sample trajectories of PAR model with Gaussian additive noise and additive outliers related to the Cases 1b and 2b considered in the simulation study.}
      \label{traj3}
\end{figure}

As the second example of the model with additive outliers we consider the most extreme case, namely, when the additive noise $\{Z_t\}$ in Eq. (\ref{model1}) is a sequence of i.i.d. random variables being a sum of two independent sequences: Gaussian additive noise and additive outliers. More precisely, we assume that for each $t$ 
\begin{eqnarray}
Z_t=Z_t^{(1)}+Z_t^{(2)}, 
\end{eqnarray}
where the sequences $\{Z_t^{(1)}\}$ and $\{Z_t^{(2)}\}$ are independent. Both of them constitute sequences of i.i.d. random variables. Moreover, for each $t$ the random variable $Z_t^{(1)}\sim N(0,0.2)$ and $Z_t^{(2)}$ is a discrete random variable such that $\mathbb{P}(Z_t^{(2)}=10)=\mathbb{P}(Z_t^{(2)}=-10)=0.003$ and $\mathbb{P}(Z_t^{(2)}=0)=0.994$.  Taking such values of the parameters of both sequences' distribution we  ensure the same level of variance of the additive noise as it was considered in case of the Gaussian additive noise (see Section \ref{simul1}) and additive outliers discussed above.  In the further analysis the  cases considered in this part are called Case 1b and Case 2b in order to notice that they correspond to Case 1 and Case 2, respectively, considered in Section \ref{simul1}.   In Fig. \ref{traj3} we demonstrate the exemplary trajectories of the considered model corresponding to Case 1b and Case 2b. 

\begin{figure}
\centering
    \includegraphics[width=\textwidth]{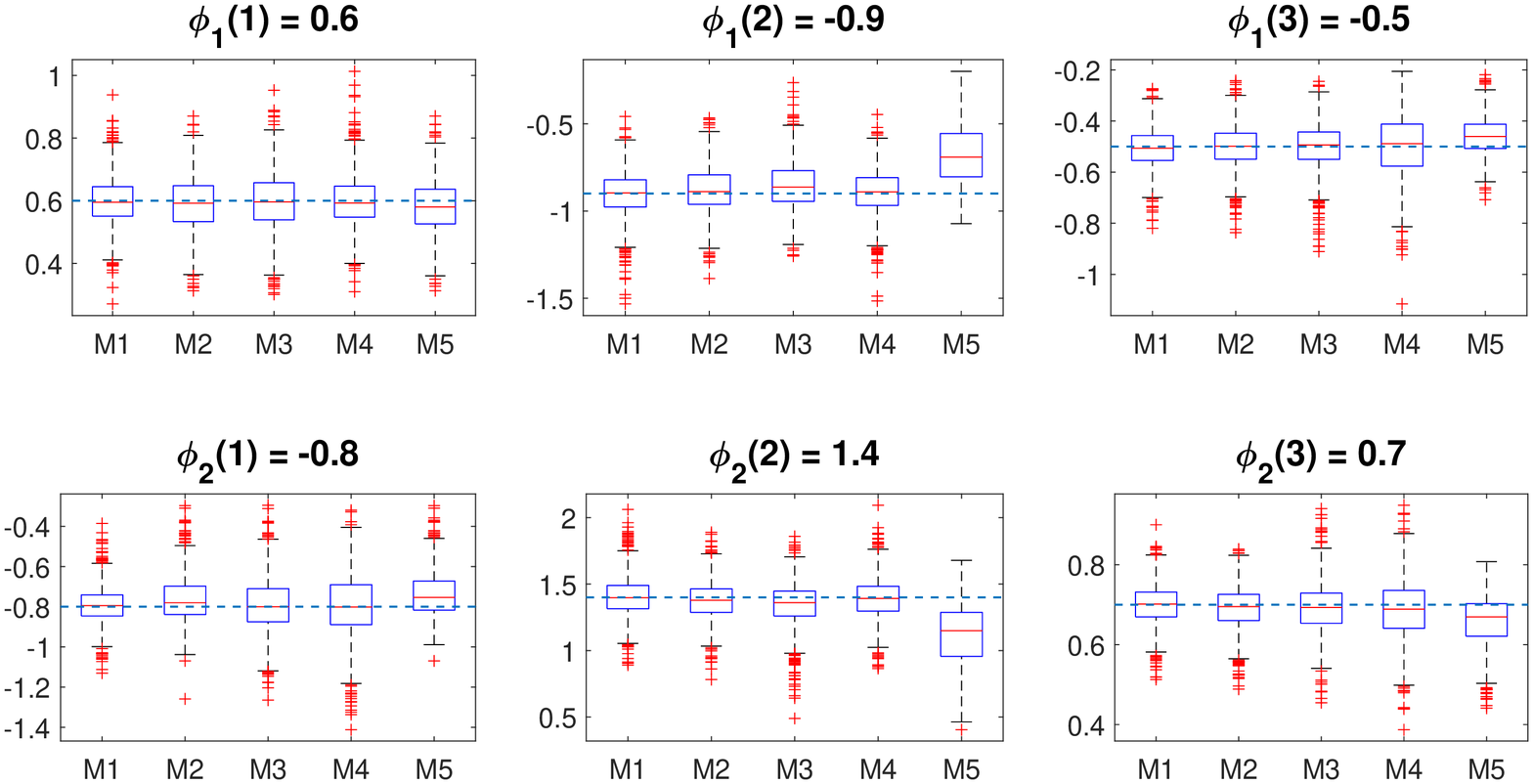}
      \caption{Boxplots of estimated values in Case 1b ($\phi_2(1) = -0.8$, $NT = 240$) for all considered methods.}
      \label{mixbp1}
\end{figure}

\begin{figure}
\centering
    \includegraphics[width=\textwidth]{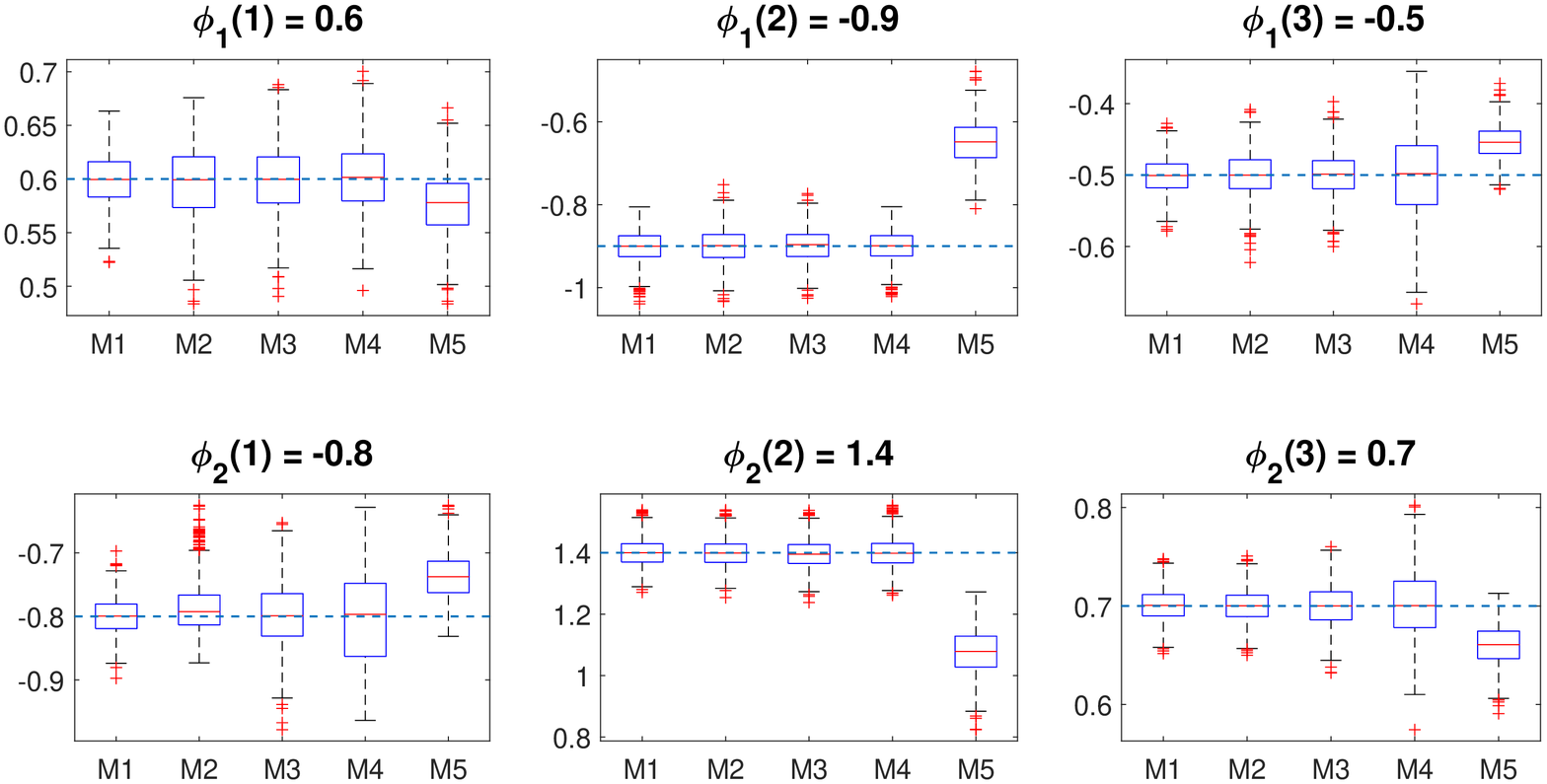}
      \caption{Boxplots of estimated values in Case 2b ($\phi_2(1) = -0.8$, $NT = 2400$) for all considered methods.}
      \label{mixbp2}
\end{figure}

The results for the Case 1b are presented in Fig. \ref{mixbp1} and Tab. \ref{mixtab1}. Moreover, the results for the Case 2b are given in Fig. \ref{mixbp2} and Tab. \ref{mixtab2}. In general, the behavior observed here can be described similarly as for Cases 1a and 2a. Once again, the advantage of proposed methods above the classical algorithm M5 is clear and becomes more visible for longer samples. In this part, just like in the previous one, the least average mean squared errors were obtained for the method M1.

\begin{table}
\centering
\begin{tabular}{| c | c | c | c | c | c | c |c|}
\hline
method & $\phi_1(1)=0.6$ & $\phi_1(2)=-0.9$ & $\phi_1(3)=-0.5$ & $\phi_2(1)=-0.8$ & $\phi_2(2)=1.4$ & $\phi_2(3)=0.7$ & average\\
\hline

M1 & \textbf{0.0060} & 0.0172 & \textbf{0.0064} & \textbf{0.0084}  & 0.0222 &\textbf{0.0027} &  \textbf{0.0105} \\ \hline
M2 & 0.0077 & 0.0171 & 0.0070 & 0.0151  &\textbf{0.0216}  & 0.0030 &  0.0119 \\ \hline
M3 & 0.0090 & 0.0206 & 0.0081 & 0.0196  & 0.0328 &0.0040  & 0.0157  \\ \hline
M4 &0.0073  &\textbf{0.0165}  &0.0144  &0.0241   & 0.0241 & 0.0057  & 0.0154  \\ \hline
M5 & 0.0077 & 0.0784 & 0.0066 & 0.0160  & 0.1295 & 0.0053 & 0.0406  \\ \hline

\end{tabular}
\caption{Parameter-wise and average mean squared errors obtained in Case 1b ($\phi_2(1)=-0.8$, $NT=240$) for all considered methods.}
\label{mixtab1}
\end{table}

\begin{table}
\centering
\begin{tabular}{| c | c | c | c | c | c | c |c|}
\hline
method & $\phi_1(1)=0.6$ & $\phi_1(2)=-0.9$ & $\phi_1(3)=-0.5$ & $\phi_2(1)=-0.8$ & $\phi_2(2)=1.4$ & $\phi_2(3)=0.7$ & average\\
\hline

M1 & \textbf{0.0005} & \textbf{0.0014} & \textbf{0.0006} & \textbf{0.0008}  & \textbf{0.0020} & \textbf{0.0003} & \textbf{0.0009}  \\ \hline
M2 & 0.0011 & 0.0016 & 0.0009 & 0.0018  & \textbf{0.0020} & \textbf{0.0003} & 0.0013  \\ \hline
M3 & 0.0010 & 0.0015 & 0.0009 & 0.0024  & 0.0021 & 0.0004 &  0.0014 \\ \hline
M4 & 0.0010 & \textbf{0.0014} & 0.0035 & 0.0048  & 0.0022 & 0.0011 & 0.0023  \\ \hline
M5 & 0.0014 & 0.0656 & 0.0026  & 0.0052  &  0.1099 & 0.0020 & 0.0311  \\ \hline

\end{tabular}
\caption{Parameter-wise and average mean squared errors obtained in Case 2b ($\phi_2(1)=-0.8$, $NT=2400$) for all considered methods.}
\label{mixtab2}
\end{table}
\section{Testing procedure}\label{testing_proc}
In this section, we present the possible application  of the proposed estimation techniques. More precisely, here we demonstrate the usefulness of the introduced algorithms for testing if  the real dataset corresponds to the pure PAR time series versus the noise-corrupted PAR model. This issue is extremely important, especially in real applications. The knowledge about the model structure can help to select appropriate tools for further data analysis.  In the $\mathcal{H}_0$ hypothesis,  we assume that the corresponding model is PAR time series with given parameters, while in the $\mathcal{H}_1$ hypothesis, the same model with additive noise is considered. In practice, we do not know the parameters of the corresponding model, however, according to the methodology presented in \cite{ZULAWINSKI2023115131}, one may identify them. Thus, in the testing procedure we assume the following
\begin{eqnarray}
    \mathcal{H}_0:~~\text{dataset corresponds to the pure PAR model with given parameters}\\
    \mathcal{H}_1:~~\text{dataset corresponds to the noise-corrupted PAR model.} \nonumber
\end{eqnarray}
In the presented testing procedure, the test statistic is the estimator of $\sigma^2_Z$, namely $\hat{\sigma}_Z^2$. Thus, the $\mathcal{H}_0$ hypothesis corresponds to the case $\sigma_Z^2=0$, while the $\mathcal{H}_1$ - to the case $\sigma_Z^2>0$. Since in this paper we consider four estimation methods, namely M1-M4, thus four test statistics are analyzed in the presented simulation study. 

The testing procedure is described in general case  assuming there is known method for estimating the parameters of noise-corrupted PAR model. We proceed as follows:
\begin{itemize}
    \item For a given dataset $x_1,x_2,\cdots,x_n$,  identify the period $T$ and order $p$ of the PAR model e.g. using the methodology presented in \cite{ZULAWINSKI2023115131}.
    \item Estimate the parameters 
    $\{\phi_i(t),\ldots,i=1,2\ldots,p\}$, $t=1,2,\ldots,T$, $\sigma^2_{\xi}$, and $\sigma_{Z}^2$ of the noise-corrupted PAR model by using the selected method presented in this paper.  
    \item Simulate $M$ Monte Carlo trajectories of length $n$ of the pure PAR model with the estimated parameters $\{\phi_i(t)\}$  and $\sigma^2_{\xi}$. In this paper, we assume Gaussian distribution of the innovations $\{\xi_t\}$.
    \item For each simulated trajectory, estimate the $\sigma_Z^2$ using the selected method.
    \item To construct the acceptance region of the test at the significance level $\alpha$, calculate the empirical quantiles at levels $\alpha/2$ and $1-\alpha/2$ (denoted as $Q_{\alpha/2}(M)$ and $Q_{1-\alpha/2}(M) $, respectively) from the estimated values of  $\hat{\sigma}_{Z}^2$ for $M$ simulated trajectories. The acceptance region is as follows:
    \begin{equation}
\label{region}
\left[Q_{\alpha/2}(M),Q_{1-\alpha/2}(M)\right]. 
\end{equation} 
\item The final step is to check if the value of the test statistic, i.e. $\hat{\sigma}^2_Z$ obtained for the analyzed dataset falls into the acceptance region. If $\hat{\sigma}^2_Z\in \left[Q_{\alpha/2}(M),Q_{1-\alpha/2}(M)\right]$, we conclude that the test does not reject the $\mathcal{H}_0$ hypothesis at significance level $\alpha$, otherwise we reject the hypothesis. 
\end{itemize}

Let us note that the test statistic considered in the presented procedure is  $\hat{\sigma}_Z^2$ which in practice should not take negative values. Thus, in  practical applications we reject the $\mathcal{H}_0$ hypothesis when the value of the test statistic is higher than the appropriate quantile and in consequence we analyze a one-sided acceptance region $(-\infty, Q_{1-\alpha}(M)]$. In the presented below simulation study, we also apply this approach.

\begin{figure}
\centering
    \includegraphics[width=0.4\textwidth]{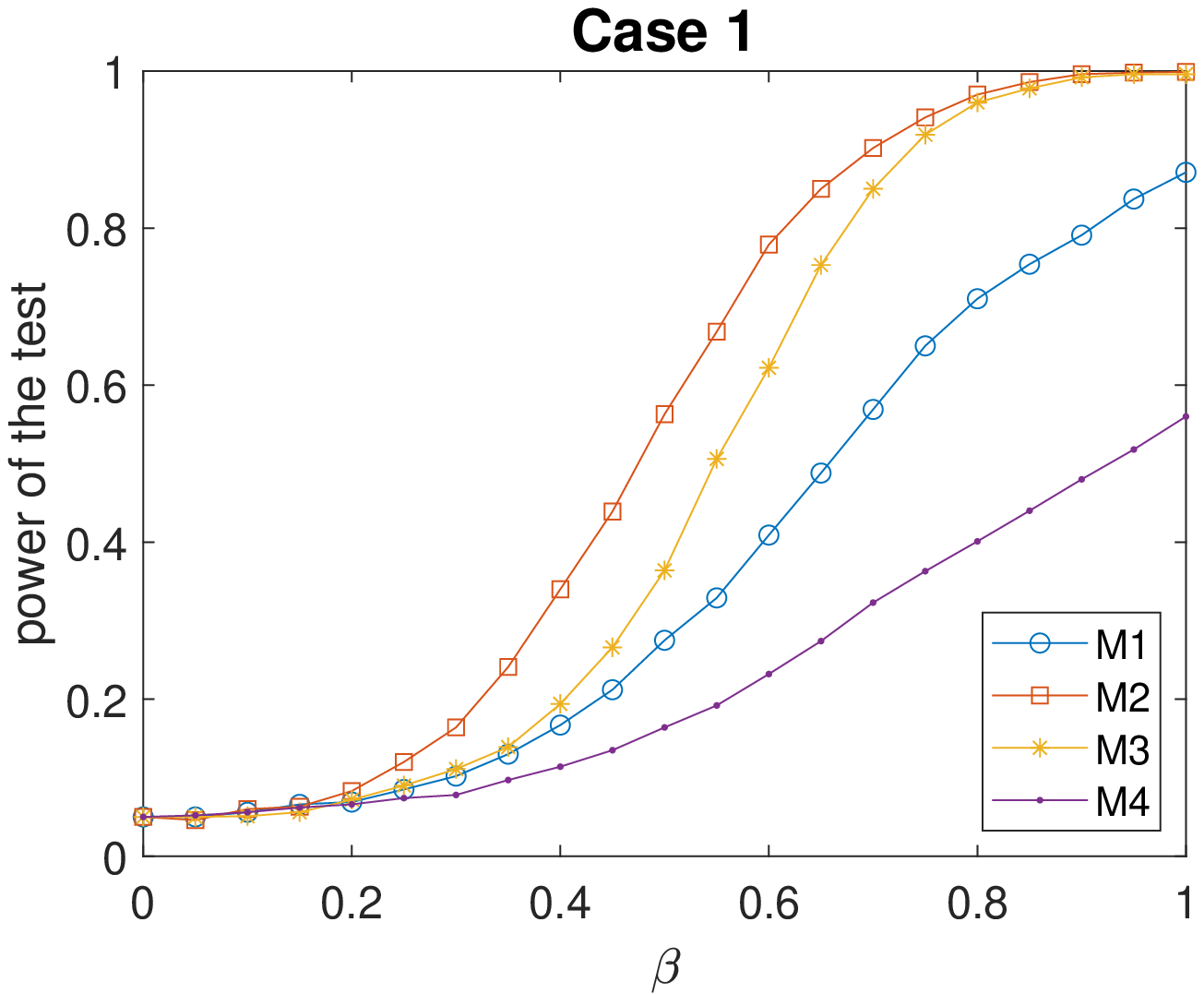}\includegraphics[width=0.4\textwidth]{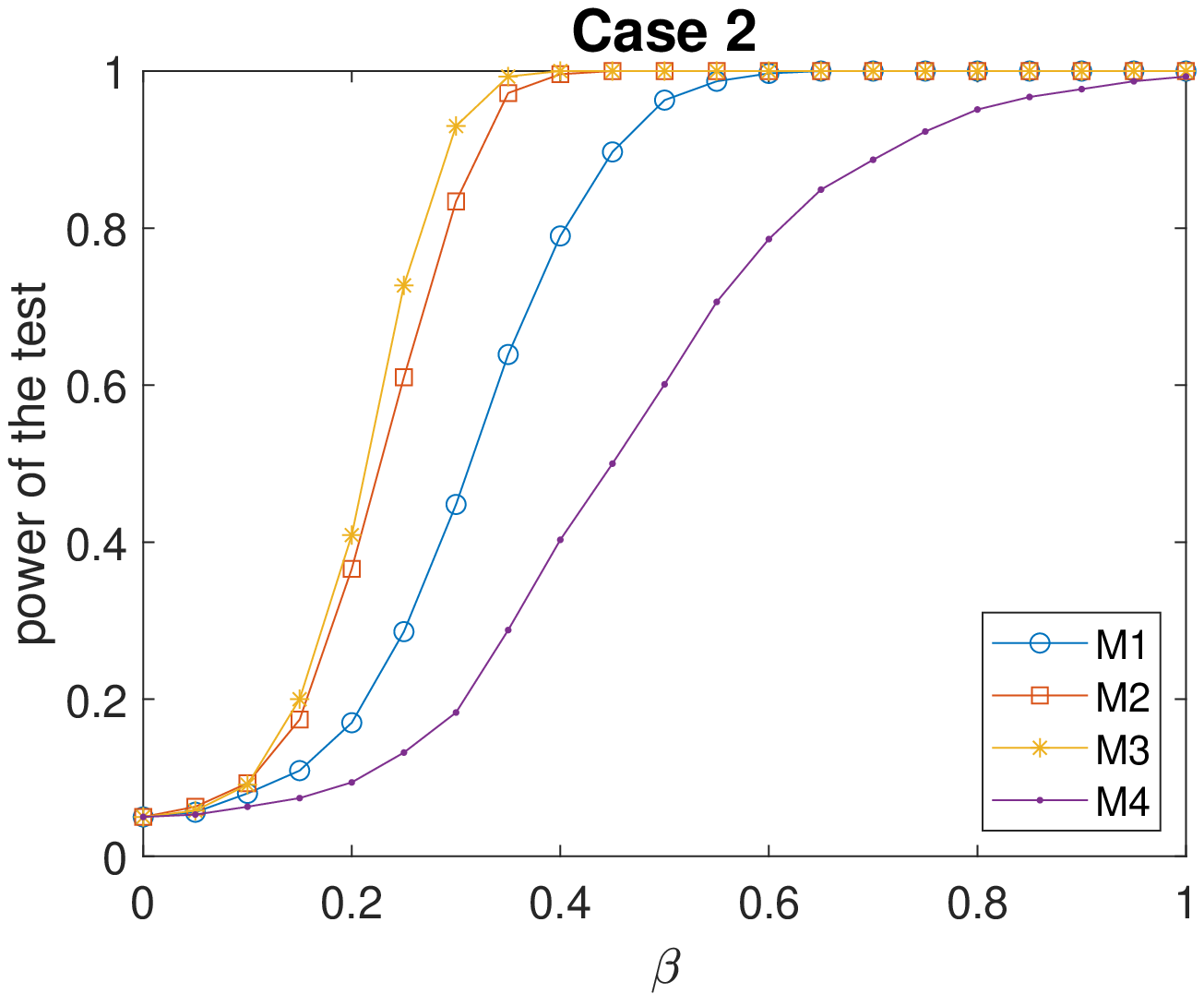}\\
      \includegraphics[width=0.4\textwidth]{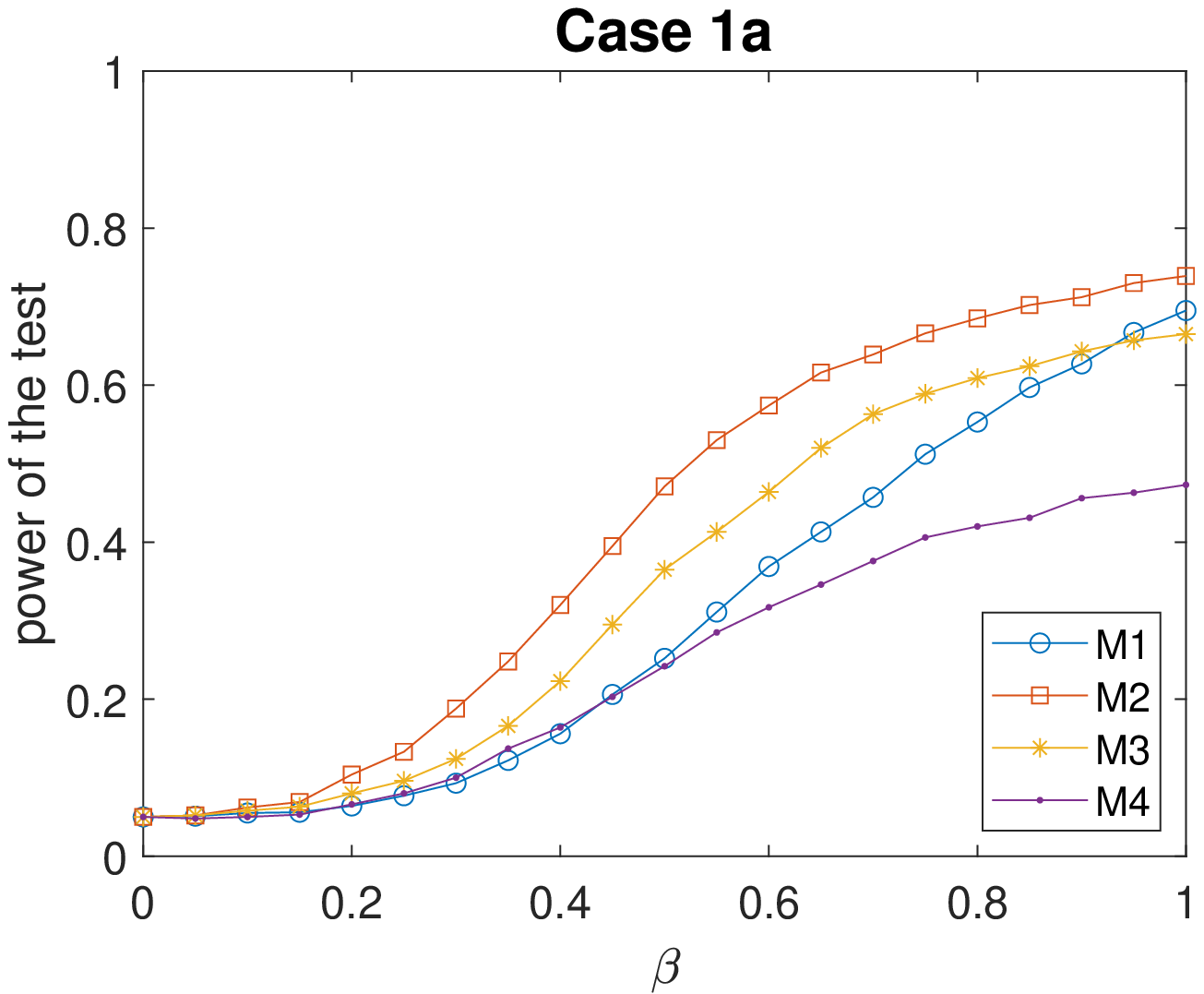}\includegraphics[width=0.4\textwidth]{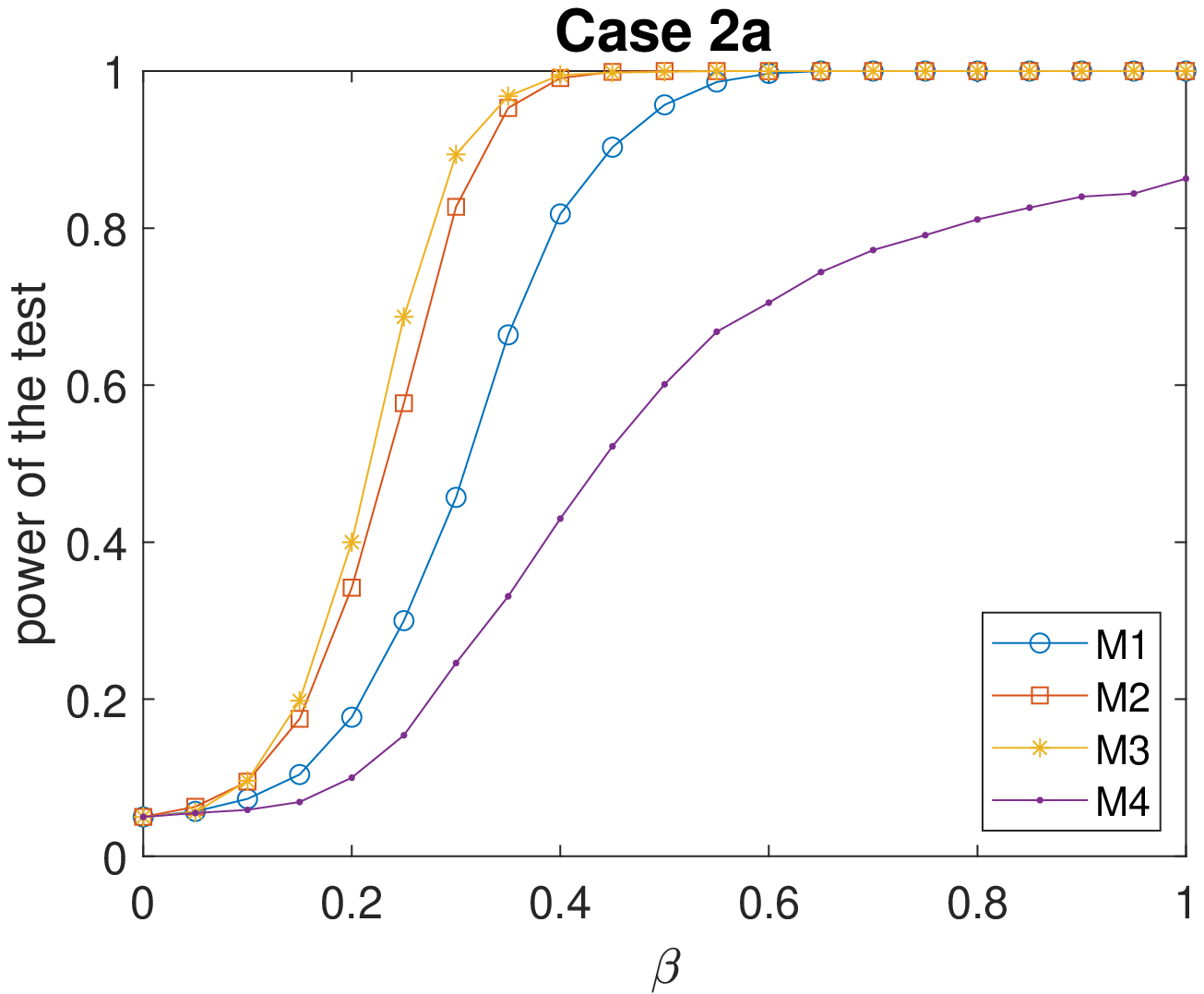}\\
        \includegraphics[width=0.4\textwidth]{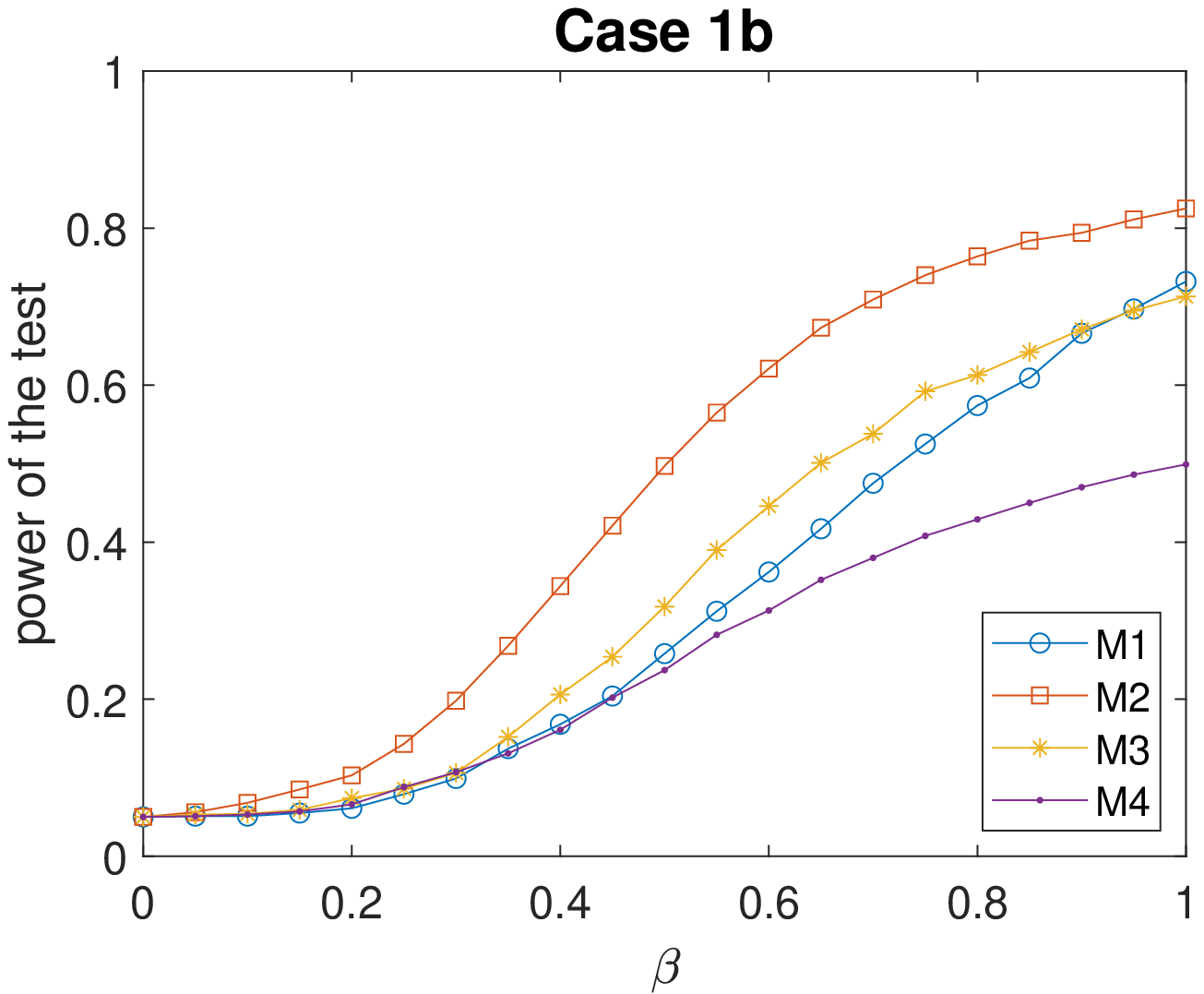}\includegraphics[width=0.4\textwidth]{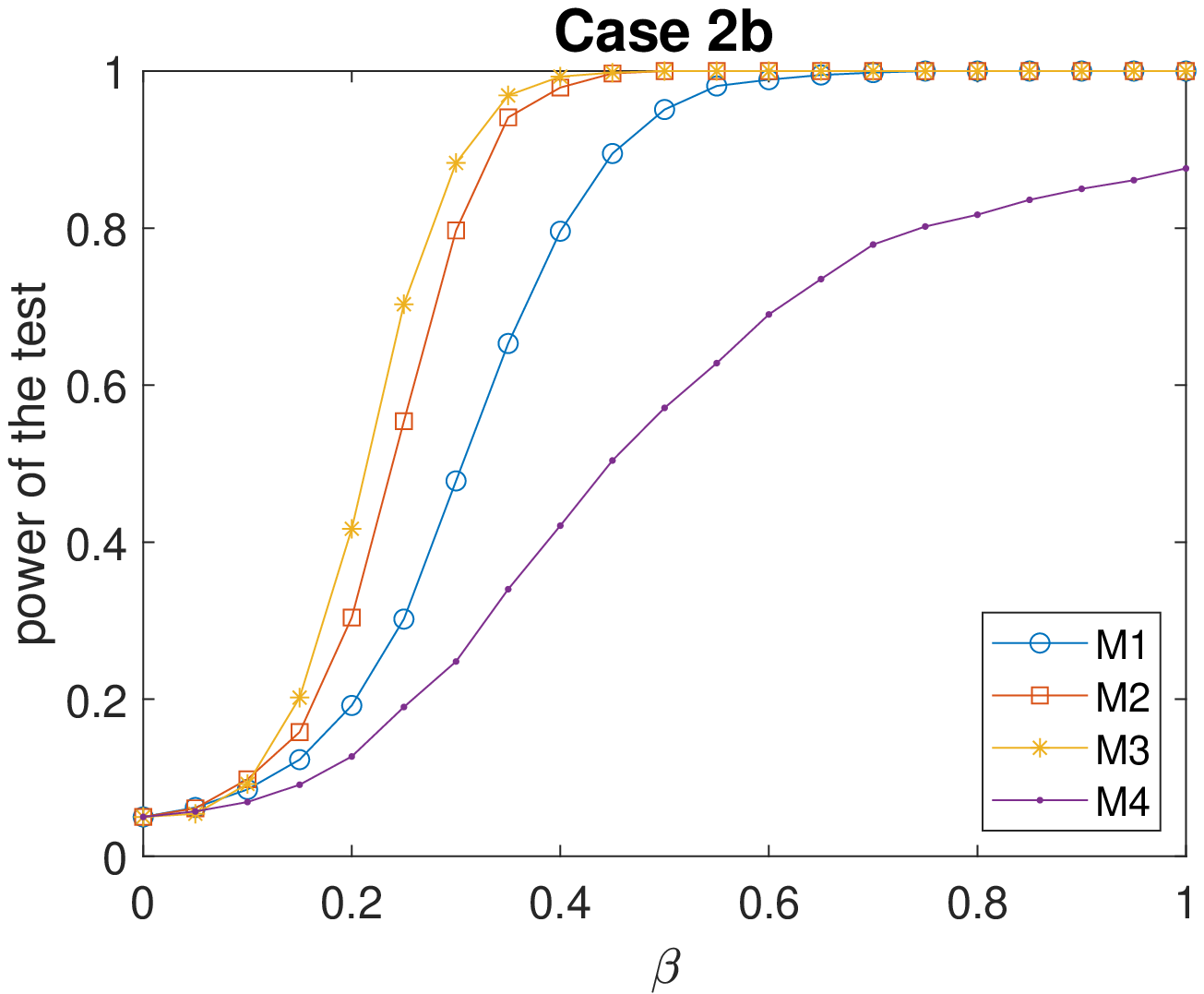}\\
      \caption{Power of the testing procedure (based on M1-M4 methods) if the dataset corresponds to the pure PAR model with parameters defined in Case 1 and Case 2 for $\alpha=5\%$. The analyzed cases correspond to three different types of additive noise distribution (each row corresponds to different distribution of the additive noise) and two different trajectory lengths (each column corresponds to different trajectory length). }
      \label{new}
\end{figure}

\begin{figure}
\centering
    \includegraphics[width=0.4\textwidth]{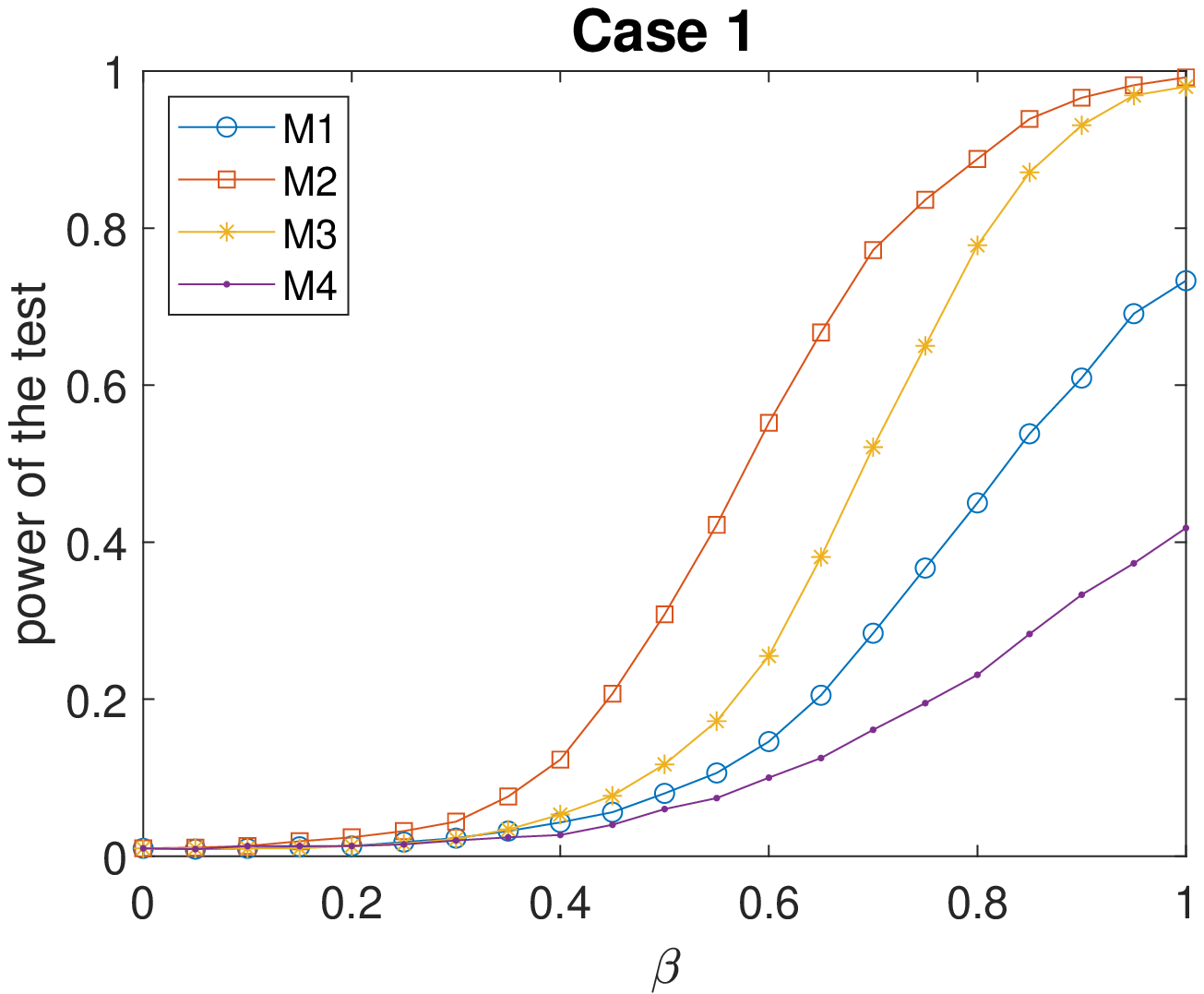}\includegraphics[width=0.4\textwidth]{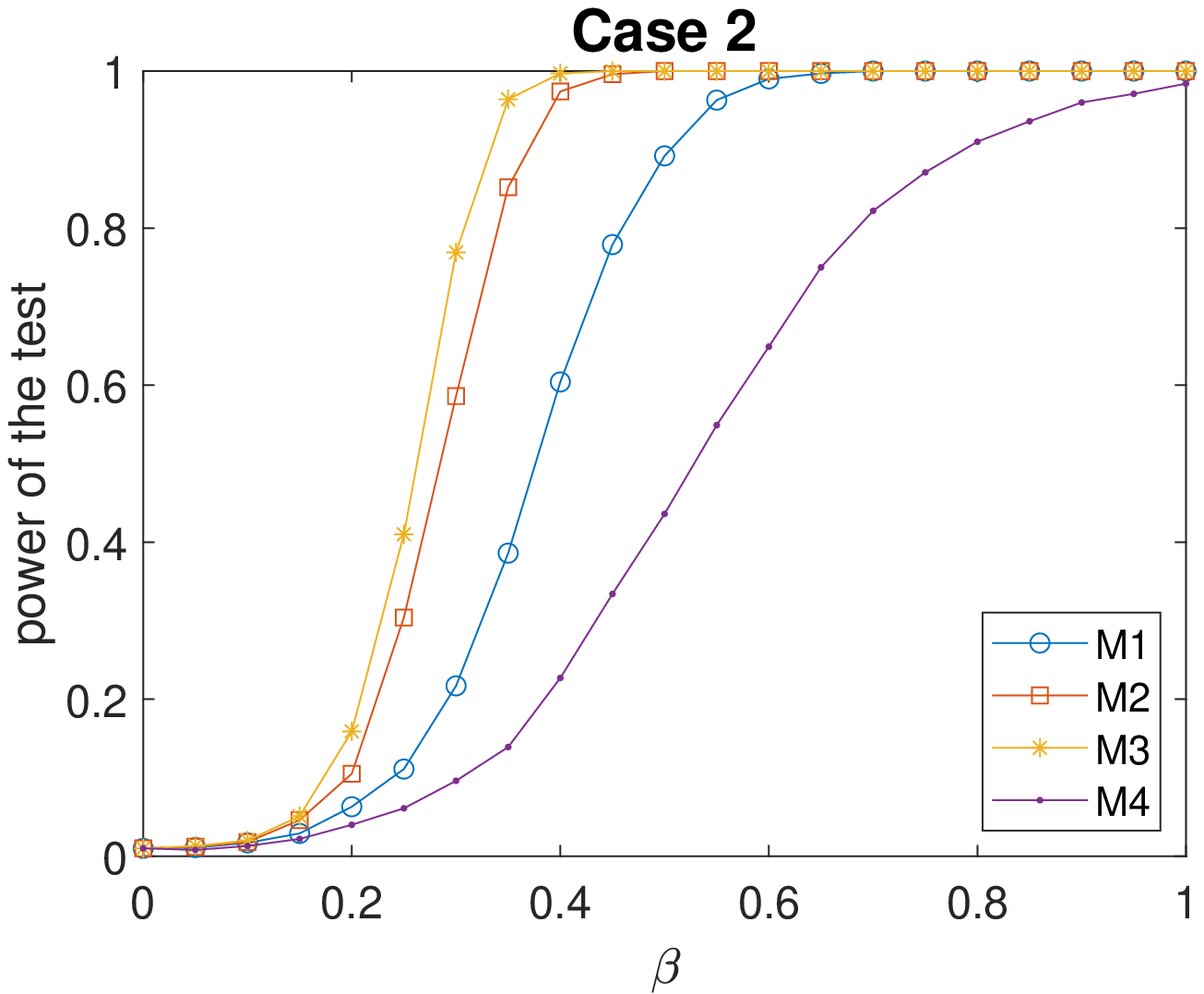}\\
      \includegraphics[width=0.4\textwidth]{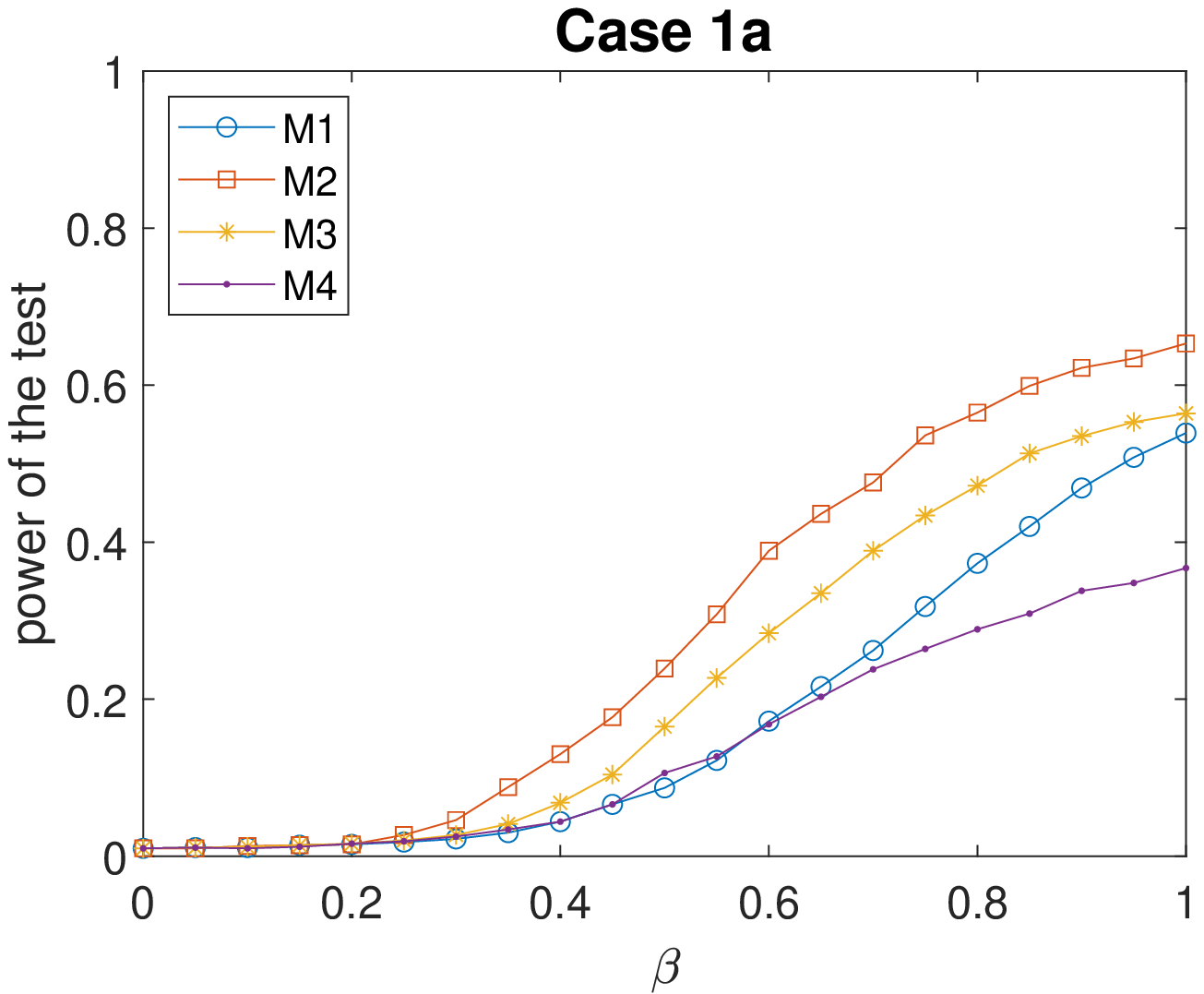}\includegraphics[width=0.4\textwidth]{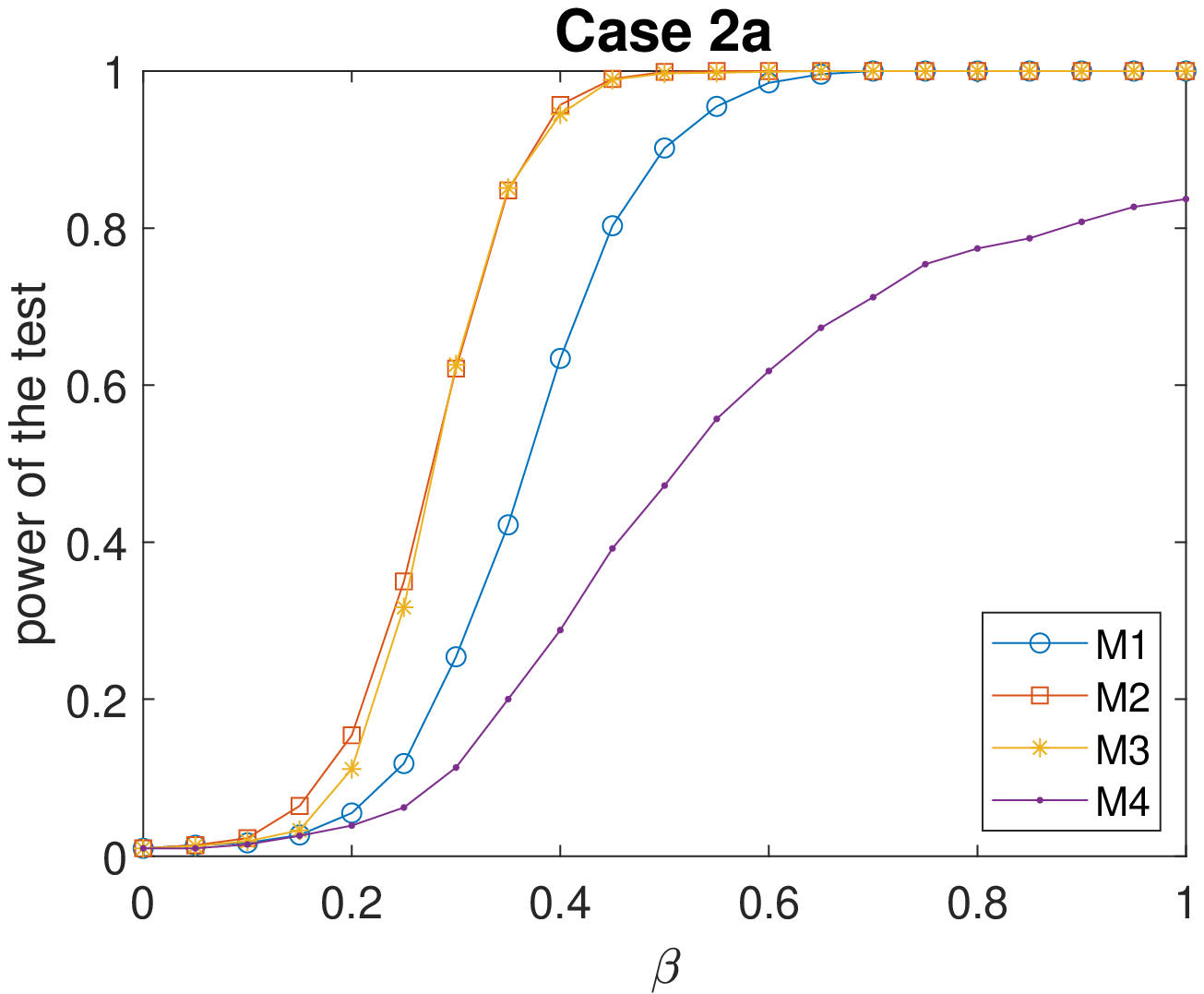}\\
        \includegraphics[width=0.4\textwidth]{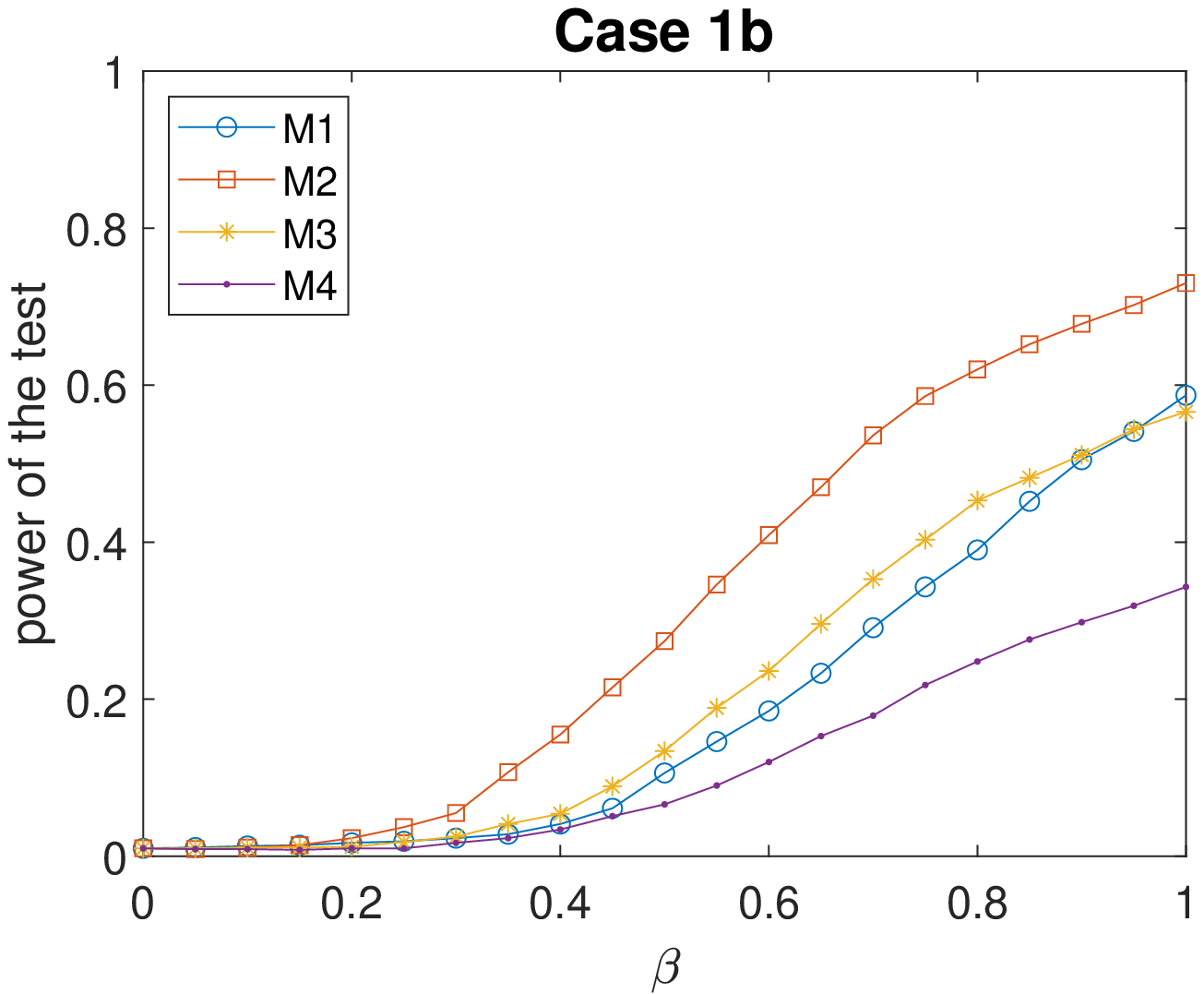}\includegraphics[width=0.4\textwidth]{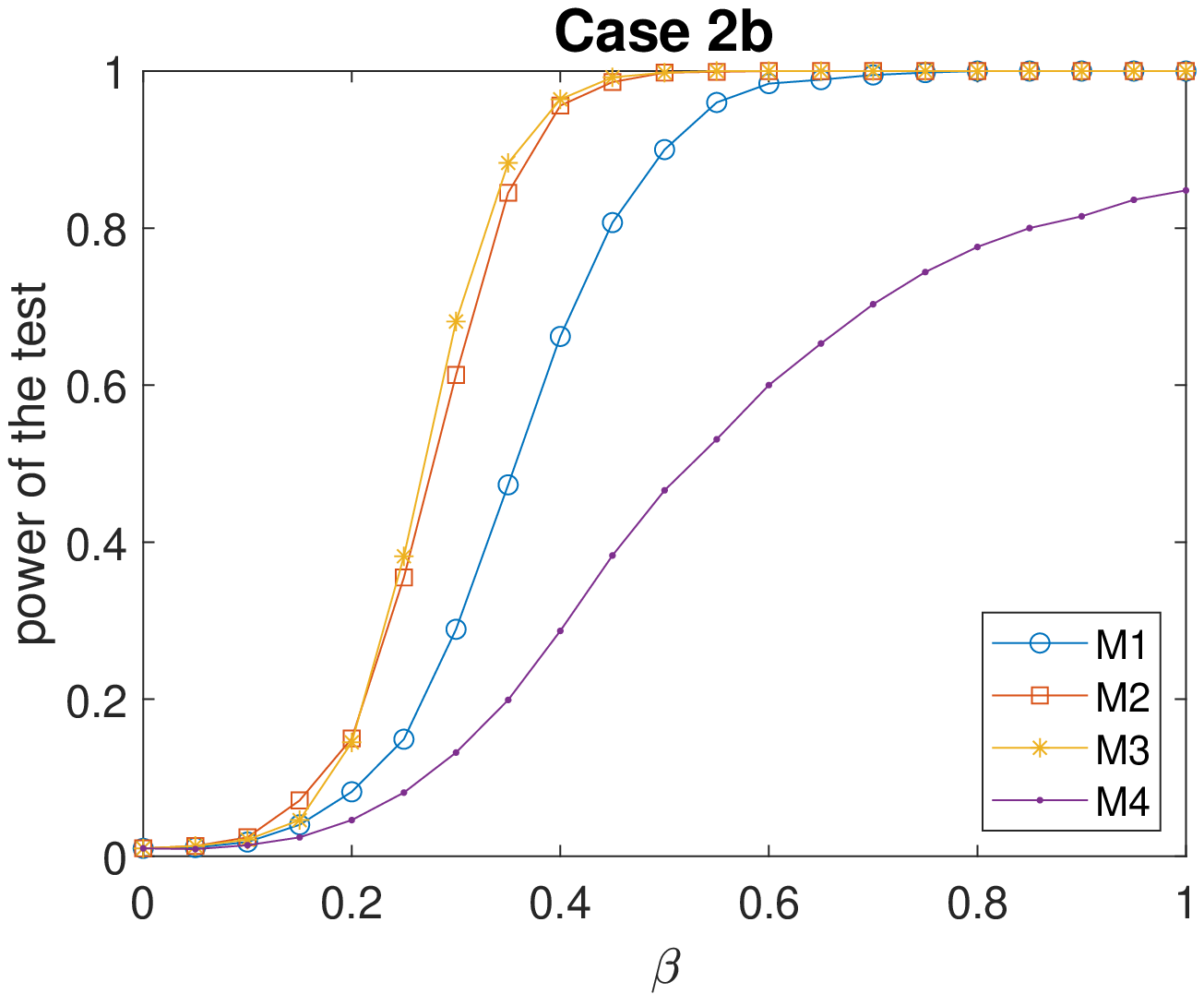}\\
      \caption{Power of the testing procedure (based on M1-M4 methods) if the dataset corresponds to the pure PAR model with parameters defined in Case 1 and Case 2 for $\alpha=1\%$. The analyzed cases correspond to three different types of additive noise distribution (each row corresponds to different distribution of the additive noise) and two different trajectory lengths (each column corresponds to different trajectory length). }
      \label{new_099}
\end{figure}

To demonstrate the efficiency of the proposed testing procedure based on M1-M4 methods, in Figs. \ref{new} and \ref{new_099} we present the power of the test for two significance levels $\alpha=5\%$ and $\alpha=1\%$ for PAR model with the parameters defined in Case 1 and Case 2 (see Section \ref{simul1}). Separately we analyze different distributions of the additive noise corresponding to the cases  Case 1, Case 2 (see Section \ref{simul1}), Case 1a, Case 1b, Case 2a, and Case 2b (see Section \ref{sec_outl}). Similar as in the simulation studies presented in Sections   \ref{simul1} and \ref{sec_outl}, we assume  $\sigma_{\xi}^2=1$. In the $\mathcal{H}_1$ hypothesis, we consider the cases when standard deviation of the additive noise is $\beta\sigma_Z$, where $\beta\in [0,0.05,0.1,\ldots,1]$ and  $\sigma_Z^2=0.8$ is the variance of the additive noise taken in the simulation studies presented in Sections \ref{simul1} and \ref{sec_outl}. We note, $\beta=0$ corresponds to the case with pure PAR model ($\mathcal{H}_0$ hypothesis).  The number of Monte Carlo simulations used to construct the acceptance region under $\mathcal{H}_0$ hypothesis is $M=1000$. The power of the test is calculated based on $M=1000$ trajectories corresponding to the $\mathcal{H}_1$ hypothesis. 

It is clearly seen that the M2 and M3 methods outperform other considered algorithms; however, for shorter trajectories (left panels of Figs. \ref{new} and \ref{new_099}), the M2 method gives the best results. For longer trajectories (right panels of Figs. \ref{new} and \ref{new_099}), both algorithms return similar power of the test. Let us note that in the case of Gaussian distributed additive noise (Case 1 and Case 2), the power of the test based on M2 and M3 reaches $1$ for smaller values of $\beta$ than for other considered cases of additive noise distribution. This effect is especially visible for shorter trajectories (left panels). 

\section{Application to the real data from air quality area}\label{real}
In this section, we demonstrate the application of the proposed testing methodology for real datasets.  The time series considered correspond to the daily mean pollutant particulate matter with diameter smaller than $10$ mm (PM$_{10}$), measured hourly in $\mu g/m^3$ and collected at the station located in the Great Vitória Region GVR-ES, Brazil, at the Automatic Air Quality Monitoring Network (RAMQAr). The complete description of the data can be found, for example, in \cite{doi:10.1080/03610926.2018.1533970}, where the authors analyzed the daily mean particulate matter concentration  from January 1, 2014 to December 29, 2015. In the mentioned bibliography position, it was claimed that the data correspond to the PAR time series with additive outliers. Thus, robust versions of the Yule-Walker method were proposed for the estimation of parameters. The air pollution data from the same region was also analyzed in various research papers in which similar models were proposed, see e.g. \cite{7362771,SARNAGLIA20102168,REISEN2019842}  and the references therein.

\begin{figure}
    \centering
    \includegraphics[width = 0.4\textwidth]{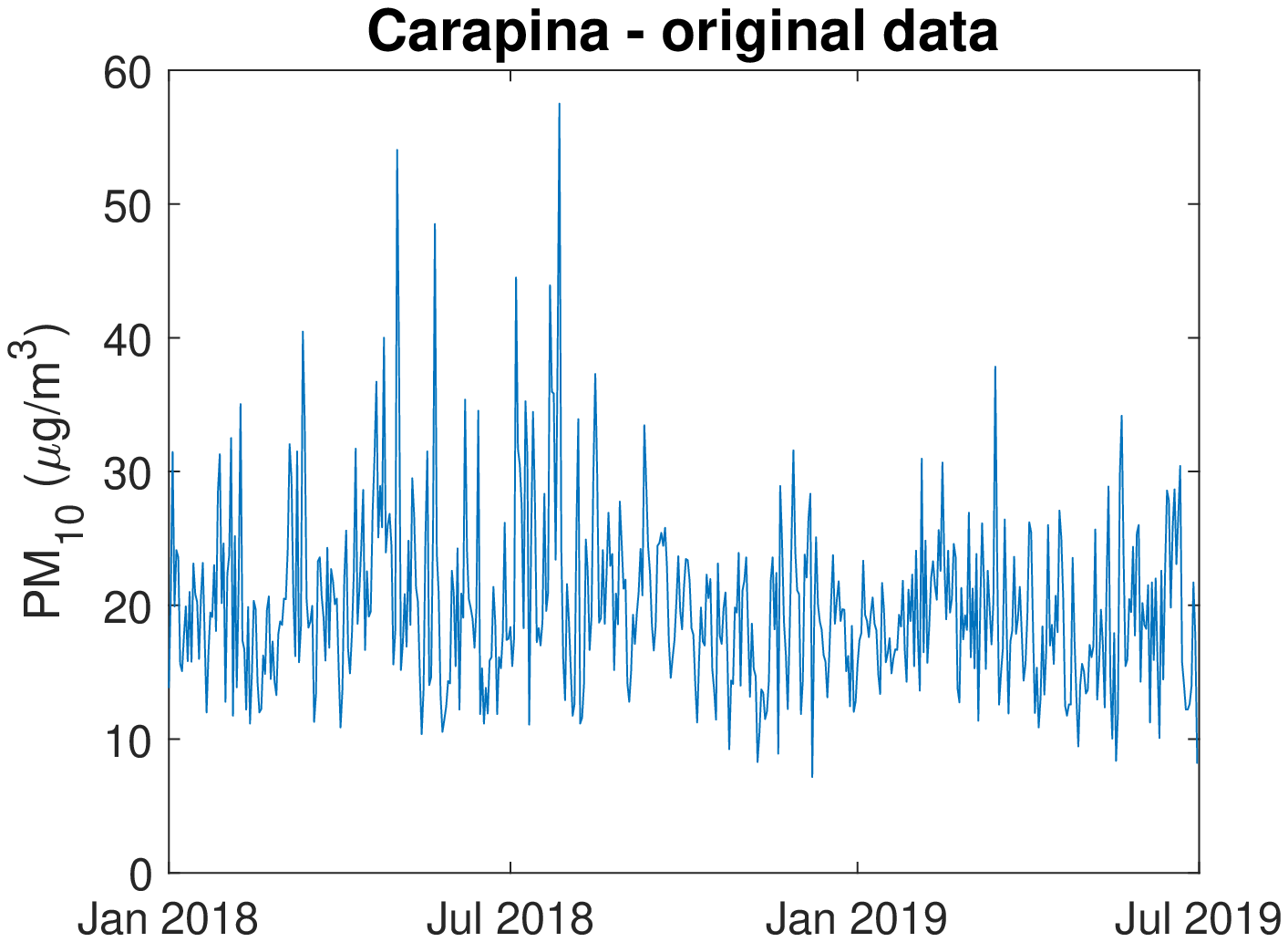}
        \includegraphics[width = 0.4\textwidth]{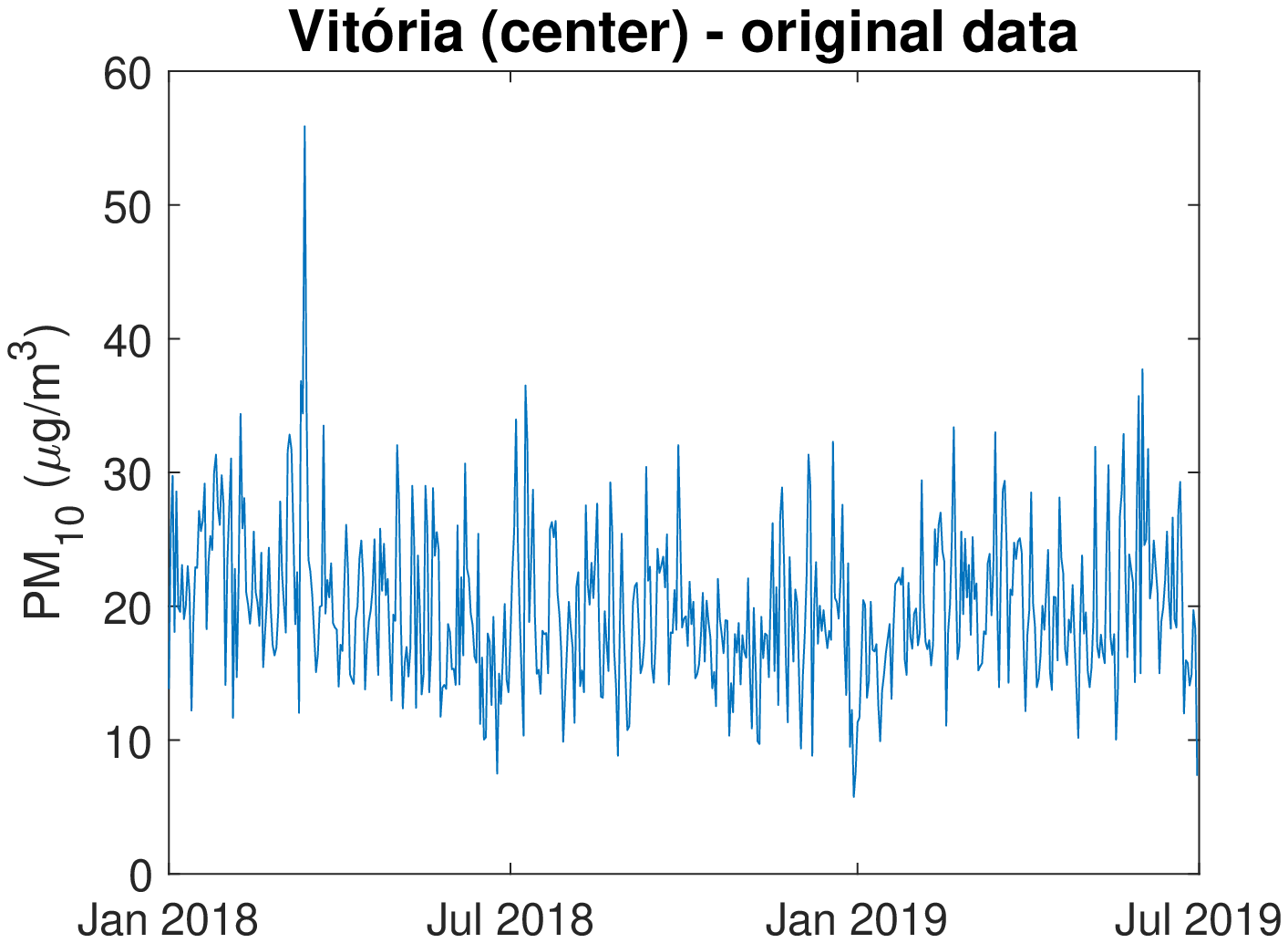}
            \includegraphics[width = 0.4\textwidth]{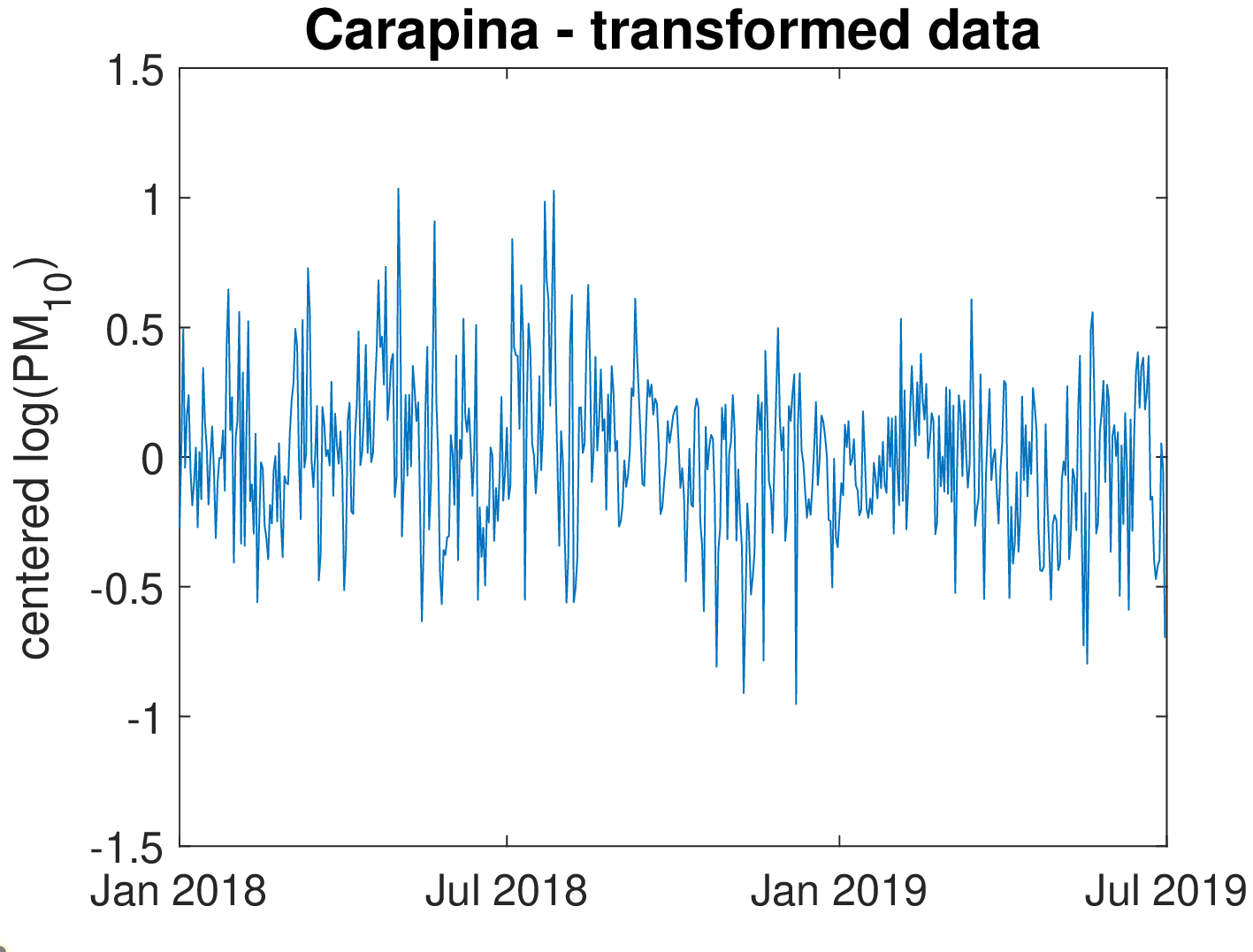}
                \includegraphics[width = 0.4\textwidth]{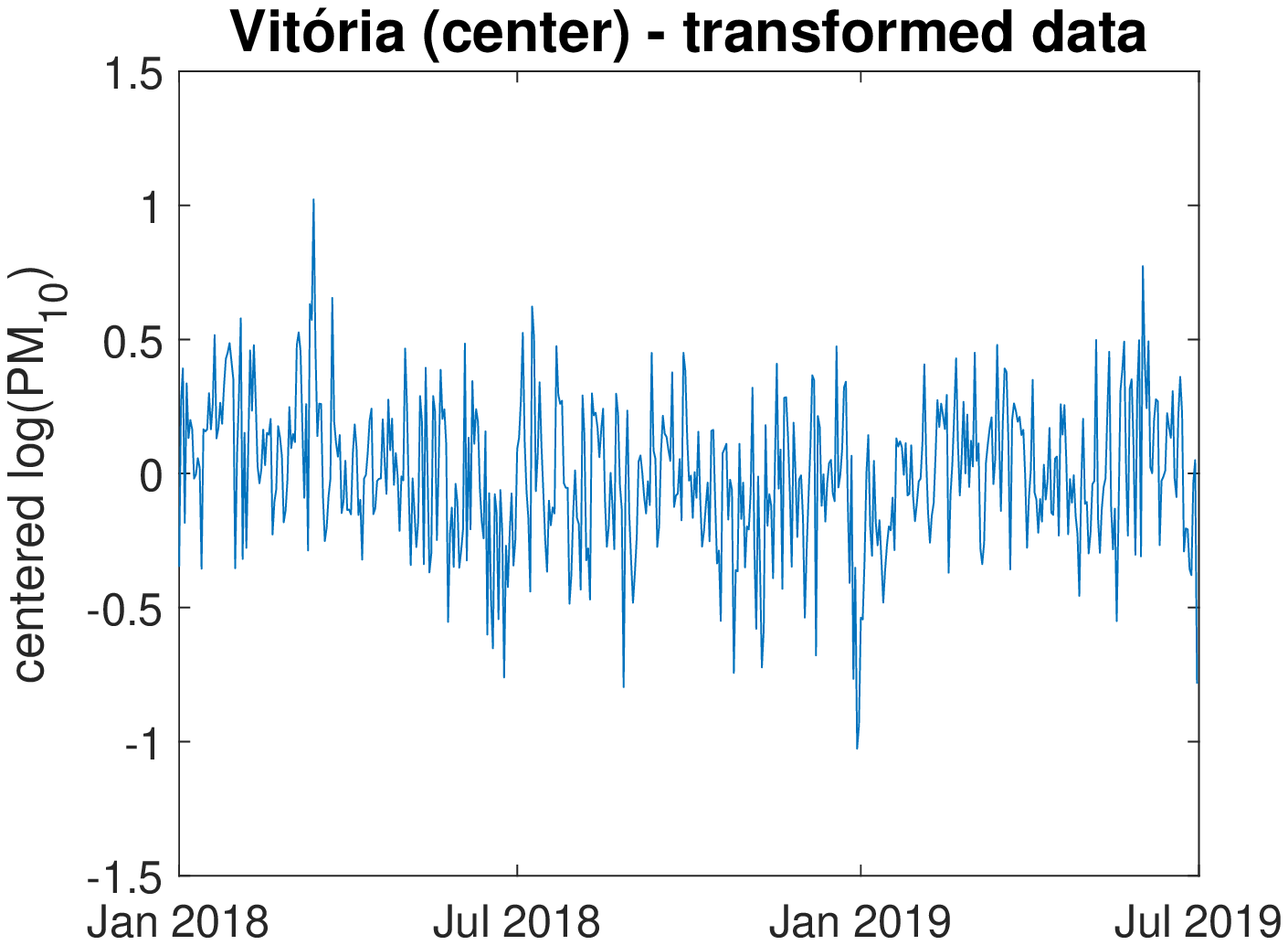}
    \caption{Two analyzed datasets --- daily average PM$_{10}$ measured in Carapina and Vitória (center) stations --- in the original (top panel) and transformed form (bottom panel).}
    \label{fig_real}
\end{figure}

In this paper, for illustration of the testing procedure, we analyze the PM$_{10}$ dataset for the period January 1, 2018 - June 30, 2019 from two monitoring stations: Carapina and Vitória (center). The real datasets are presented in Fig. \ref{fig_real} (top panel). Similarly as in the papers mentioned above, the data were pre-processed to eliminate the skewness and some evidence of time-varying variance. First, the natural logarithm was applied. Then, as in \cite{parma_ao3}, the data were centered by subtracting the Huber location $M$-estimator for each period. In Fig. \ref{fig_real} (bottom panel), we present the transformed data that are further taken for analysis. In Fig. \ref{fig_real}, especially in the top panels, one can see large peaks of the PM$_{10}$ concentration that may be viewed here as outliers. The same conclusion was drawn in \cite{doi:10.1080/03610926.2018.1533970}, where the authors indicated the importance of additive outliers in the estimation results using classical approaches (i.e., classical Yule-Walker method). In this paper, we continue this research and provide the testing procedure described in Section \ref{testing_proc} for the PAR model with $T=7$ and $p=4$. The used period $T$ and the order $p$ were identified by the authors \cite{doi:10.1080/03610926.2018.1533970} for similar datasets. 

Since the simulated data analysis clearly indicated M2 and M3 as the most efficient methods, here we present the results only for those methods. In Tab. \ref{real_table}, we demonstrate the test statistic values obtained for the real datasets and the acceptance regions obtained based on $M=1000$ simulations of the tested models for two values of significance level, namely $\alpha=5\%$ and $\alpha=1\%$. As mentioned, the model corresponding to $\mathcal{H}_0$ hypothesis is a PAR time series with Gaussian innovations. Our results clearly confirm the results presented in the above-mentioned articles. For both significance levels, we reject the $\mathcal{H}_0$ hypothesis of pure PAR time series. Thus, for further analysis, it is reasonable to apply the dedicated methods for estimation of the model parameters, as was performed in \cite{doi:10.1080/03610926.2018.1533970}, where the authors proposed to apply the robust versions of the Yule-Walker approach. 

\newpage
\begin{table}
\centering
\begin{tabular}{| c |c| c | c | c |}
\hline
dataset & method & test statistic value $\hat{\sigma}_Z^2$ & acceptance region, $\alpha=5\%$ & acceptance region, $\alpha=1\%$ \\ \hline
Carapina & M2 & 0.0173 & $[0,0.0084]$ & $[0,0.0102]$ \\ \hline
Carapina & M3 & 0.0216 & $[0,0.0072]$ & $[0,0.0096]$ \\ \hline
Vitória (center) & M2& 0.0129 & $[0,0.0073]$ & $[0,0.0091]$\\ \hline
Vitória (center) & M3& 0.0197 & $[0,0.0040]$ & $[0,0.0053]$\\ \hline
\end{tabular}
\caption{Test statistic values and acceptance regions obtained for both analyzed datasets using estimation methods M2 and M3.}
\label{real_table}
\end{table}

\section{Summary and discussion}
In this paper, we have discussed the problem of identification and characterization of the general noise-corrupted PAR model with finite variance. The main attention was paid to the estimation of the parameters of the considered model. We have proposed four Yule-Walker-based estimation techniques and demonstrated their efficiency for different cases of the model's parameters and three additive noise types, namely, Gaussian distributed noise and two cases of non-Gaussian distributed noise (additive outliers, and the case when the disturbances are the sum of the Gaussian noise and additive outliers). On the basis of the proposed estimation techniques, we also introduced the testing procedure for identification if the data correspond to the pure PAR model versus the noise-corrupted one.

The results of the current paper extend our previous research presented in \cite{nasza_wojtek} and \cite{ZULAWINSKI2023115131}. In the first paper, a particular case of the model was examined, namely PAR(1) while in the second one, the procedure for selection of the orders of noise-corrupted PAR model was discussed, and the technique for identification of residuals probabilistic properties is proposed.  The methodology presented in the current paper can be considered universal. It can be applied to noise-corrupted PAR models of any order $p$ and period $T$ under the assumption of finite variance distribution. Moreover, the proposed techniques are dedicated to any type of additive noise distribution under the assumption of finite variance. 

One of the main motivations of the presented research were the data from the air quality area. In the papers \cite{doi:10.1080/03610926.2018.1533970,7362771,SARNAGLIA20102168,REISEN2019842} the authors identified the existence of additive outliers in time series corresponding to particulate matter concentration. For them, this observation was a starting point for the introduction of robust estimation methods that are not sensitive for large observations observed in the data. In the current research, we went a step further and considered a more general case, i.e. the problem of the existence of a general finite-variance additive noise in the PAR model. The techniques proposed in this paper are universal and can also be applied in the case of additive outliers. Moreover, we present the application of the introduced estimation methods for testing the existence of additive noise. This issue is significant in real applications, where the appearance of additive noise may influence the selection of appropriate tools for further analysis. 

The second motivation of the current research comes from condition monitoring, where the models of real vibration might be used to develop an inverse filter to remove components related to mesh frequencies in gearbox vibrations. Analysis of the residual signal allows one to detect local damage. The lack of precision during modelling results in distortions in residuals and poor damage detection efficiency \cite{Wylomanska2014171}. The presence of additive noise in the real vibration signals is very intuitive. In addition, the existence of additive outliers in the signal is also often observed in vibrations and may significantly influence the estimation results. Therefore, the introduction of estimation methods and testing procedures dedicated to the model with general additive disturbances may improve the efficiency of the techniques used for local damage detection.   

In the presented analysis, we have discussed only a few aspects of the considered problem, such as, e.g., the influence of the additive outliers on the estimation results,  the sensitivity of the new techniques to the signal length, and comparison of the efficiency of the proposed techniques in the considered cases. However, the presented results open new areas of interest that are crucial for real data applications. 

\section*{Acknowledgements}
This work is supported by National Center of Science under Sheng2 project No. UMO-2021/40/Q/ST8/00024 "NonGauMech - New methods of processing non-stationary signals (identification, segmentation, extraction, modeling) with non-Gaussian characteristics for the purpose of monitoring complex mechanical structures". \\
The authors would like to thank prof. Valderio Reisen for sharing the real data analyzed in this paper.

\bibliography{mybibliography}

\end{document}